\documentclass[useAMS,usenatbib,epsfig]{mn2e}
\usepackage{natbib}
\usepackage{amsmath,amsfonts}
\usepackage{epsfig}

\def\reff@jnl#1{{\rm#1\/}}

\def\aj{\reff@jnl{AJ}}                  
\def\araa{\reff@jnl{ARA\&A}}            
\def\apj{\reff@jnl{ApJ}}                        
\def\apjl{\reff@jnl{ApJ}}               
\def\apjs{\reff@jnl{ApJS}}              
\def\ao{\reff@jnl{Appl.Optics}}         
\def\apss{\reff@jnl{Ap\&SS}}            
\def\aap{\reff@jnl{A\&A}}               
\def\aapr{\reff@jnl{A\&A~Rev.}}         
\def\aaps{\reff@jnl{A\&AS}}             
\def\azh{\reff@jnl{AZh}}                        
\def\baas{\reff@jnl{BAAS}}              
\def\jrasc{\reff@jnl{JRASC}}            
\def\memras{\reff@jnl{MmRAS}}           
\def\mnras{\reff@jnl{MNRAS}}            
\def\pra{\reff@jnl{Phys.Rev.A}}         
\def\prb{\reff@jnl{Phys.Rev.B}}         
\def\prc{\reff@jnl{Phys.Rev.C}}         
\def\prd{\reff@jnl{Phys.Rev.D}}         
\def\prl{\reff@jnl{Phys.Rev.Lett}}      
\def\pasp{\reff@jnl{PASP}}              
\def\pasj{\reff@jnl{PASJ}}              
\def\qjras{\reff@jnl{QJRAS}}            
\def\skytel{\reff@jnl{S\&T}}            
\def\solphys{\reff@jnl{Solar~Phys.}}    
\def\sovast{\reff@jnl{Soviet~Ast.}}     
\def\ssr{\reff@jnl{Space~Sci.Rev.}}     
\def\zap{\reff@jnl{ZAp}}                        
\def\nat{\reff@jnl{Nature}}             

\title[High Galactic latitude Observations with the COSMOSOMAS
Experiment]{Observations of the Cosmic Microwave Background
and Galactic Foregrounds at 12-17 GHz with the COSMOSOMAS Experiment}

\author[S. Fern\'andez-Cerezo et al.]{S. Fern\'andez-Cerezo$^1$, 
C. M. Guti\'errez$^1$, R. Rebolo$^{1,2}$, 
R. A. Watson$^{3,\dagger}$,  
 \newauthor  R. J. Hoyland$^1$, S. R. Hildebrandt$^1$, J. A.
 Rubi\~no-Mart\'\i n$^1$, 
 \newauthor J. F. Mac\'\i as-P\'erez$^4$, and P. Sosa
Molina$^1$ \\ \\
  $^1$Instituto de Astrof\'\i sica de Canarias, 38200 La Laguna,
  Tenerife, Spain.\\ 
  $2$ Consejo Superior de Investigaciones Cient\'\i ficas, Spain. \\
  $^3$ Jodrell Bank Observatory, University of Manchester, 
  Macclesfield, Cheshire SK11 9DL, UK.\\
 $^4$ Laboratoire de Physique Subatomique et Cosmologie, 53, av. des
 Martyrs, Grenoble, France. \\ 
  $^{\dagger}$Present address: Instituto de Astrof\'\i sica de
Canarias, 38200 La Laguna, Tenerife, Spain. }

\begin{document}
\date{Accepted ---; Received ---; in original form \today}
\pagerange{\pageref{firstpage}--\pageref{lastpage}}
\pubyear{2005}
\maketitle
\label{firstpage}

\begin{abstract}

We present the analysis of the first 18 months of data obtained with
the COSMOSOMAS experiment at the Teide Observatory (Tenerife). Three
maps have been obtained at 12.7, 14.7 and 16.3 GHz covering 9000
square degrees each with a resolution of $\sim 1$ degree and with
sensitivities 49, 59 and 115 $\mu$K per beam respectively. These data in conjuction 
with the WMAP first year maps have
revealed that the Cosmic Microwave Background (CMB) is the dominant
astronomical signal at high galatic latitude ($|b|>40^\circ$) in the
three COSMOSOMAS channels with an average amplitude of $29.7\pm 1.0$
$\mu$K (68\% c.l. not including calibration errors). This value is in
agreement with the predicted CMB signal in the COSMOSOMAS maps 
using the   best fit $\Lambda$-CDM model to the WMAP power spectrum.  
Cross correlation analysis of the 408~MHz 
map and the COSMOSOMAS data at high galactic latitudes give
values  in the range $17.0-14.4~\mu$K from 12.7 
to 16.3 GHz. Removing detected point sources in this template, reduces the amplitude of
the correlated signal to  8-9 $\mu$K. 
The mean spectral index of the correlated signal between the 408 MHz desourced  
and the COSMOSOMAS
maps is between   -3.20 and -2.94  at
$|b|$$>$40$^\circ$ which indicates that this signal is  due to synchrotron 
emission.  Cross-correlation of COSMOSOMAS data with the
DIRBE map at 100 $\mu$m shows the existence
of a common signal with  amplitude $7.4\pm 1.1$, $7.5\pm 1.1$, and $6.5\pm 2.3$
$\mu$K in the 12.7, 14.7 and 16.3 GHz COSMOSOMAS maps at
$|b|>$30$^\circ$.  Using the WMAP data we find this DIRBE correlated
signal rises from high to low frequencies 
flattening below $\sim 20$ GHz.  At higher galactic latitudes
the average amplitude of the correlated signal with
the DIRBE maps decreases slightly. 
The frequency behaviour of the COSMOSOMAS/WMAP correlated
signal with DIRBE is not compatible with the expected tendency for
thermal dust. A study of the H$\alpha$ emission maps do not support
free-free as a major contributor to that signal. Our results
provide evidence of a new galactic foreground with properties
compatible with those predicted by the spinning dust models.

\end{abstract}


\section{Introduction}

In recent years many experiments have measured Cosmic Microwave Background (CMB) anisotropies
at angular scales from several degrees to a few arcmin. The analysis of the resulting power
spectra has tested the Lambda Cold Dark Matter ($\Lambda$CDM) model and allowed an accurate
determination of cosmological parameters ($e.\,g.$ Spergel et al. 2003; Rebolo et al.  2004).
Detailed knowledge of the galactic foregrounds (synchrotron, free-free and vibrational thermal
dust emission) is required to ensure the validity of the results. The joint analysis of
different templates of galatic emission with WMAP observations (Bennett et al. 2003; Hinshaw
et al. 2003) has allowed a better assesment of the properties of these galactic foregrounds at
the frequencies where CMB searches are conducted (see also Banday et al. 2003). Apart from the
three classical diffuse components (synchrotron, free-free and vibrational dust  emission),
there is increasing evidence of an additional dust correlated foreground (Casasus et al. 2004; de
Oliveira-Costa et al. 1997, 1998, 1999, 2004; Finkbeiner et al. 2002; Kogut et al. 1996)
 which does not follow the expected thermal dust
spectrum at  frequencies below $\nu\la$40 GHz.  The physical nature of such a component is
intriguing (Lagache 2003; Leitch et al. 1997; Lim et al. 1996; Mukherjee et al. 2001; Banday
et al. 2003). Although it was initially ascribed to free-free emission, the low level of
related H$\alpha$ emission makes unlikely this explanation. Alternatively, one of the most
attractive scenarios to explain the origin of that emission is electric dipole radiation from
spinning small dust particles as proposed by Draine \& Lazarian (1998$a,b$). The key
observational prediction of such a model is a turn-over in the spectrum at frequencies
10-20~GHz which are not covered by COBE or WMAP.

The Tenerife experiment (Guti\'errez et al. 2000) has provided observations at 10 and 15 GHz
able to test this prediction ($e.\,g.$ de Oliveira et al 2004). The COSMOSOMAS (COSMOlogical Structures
On Medium Angular Scales) experiment represents a qualitative improvement in terms of angular
resolution, sensitivity and sky coverage to study any possible new foreground in the frequency
range 10-17~GHz. This experiment, designed and built at the Instituto de Astrof\'\i sica de
Canarias, is located at an altitude of 2400 m in the Teide Observatory, Tenerife (Spain). It
consists of two similar instruments working at frequencies centred at 11 and 15 GHz named
COSMO~11 and COSMO~15 respectively.  This paper is dedicated exclusively to the analysis of
COSMO~15 data. The COSMO~11 results will be presented in a separate paper (Hildebrandt et al.
2005 in prep.). A detailed technical description of the COSMO~15 instrument has been presented
in Gallegos et al. (2001) where we also reported the first results obtained with this
experiment. Here, we present additional data taken up to January 2001 improving the
sensitivity by a factor two. We use the COSMOSOMAS data in conjuction with the first year WMAP
data and several Galactic templates to make a robust estimation of the CMB and the different
components of the galactic diffuse emission in the frequency range 12-17 GHz. 

\section{Experimental setup, observations and  data reduction}

\subsection{Instrument}

COSMO~15 uses a flat rotating circular mirror (2.5 m of diameter) 
made of aluminum canted by 5 degrees with respect to its rotation axis. The
sky radiation is reflected by that mirror and redirected to a 2.4 m off axis
parabolic antenna where it is focused and detected by a cryo-cooled HEMT
amplifier. The rotation of the mirror produces circular scans in the sky at
1~Hz. The data are sampled $\sim 220$ times per turn of the mirror. Earth
rotation allows the daily observation of a sky strip complete in right
ascension and 20 degrees width in declination. The instrument has three
independent channels (hereafter called C1, C2 and C3) working at frequencies
centred at 12.7, 14.7 and 16.3 GHz respectively with bandwidths $\sim 1$ GHz
and angular resolution $\sim 1$ degree.  The instrument is shielded to
protect against spillover. To correct for changes in the gain of the
instrument, every 30 s a constant signal at 2 K is injected.  System
temperatures are $\sim 35$ K with small differences between channels.
Table~\ref{param} summarizes the main parameters including beam sizes
estimated as explained in Section 3.2.

\begin{table}
\caption{The three channels of the COSMO~15 instrument.}
\begin{center}
\begin{tabular}{cccccc}
\hline
\hline
 Channel &Freq. & Bandwidth & T$_{sys}$ & FWHM$_{RA}$  &
 FWHM$_{Dec.}$ \\
 & (GHz) & (GHz) & (K) & (deg) & (deg) \\
\hline
C1& 12.7 & 1.17 & 38.5 & $0.92\pm 0.01$ & $1.11\pm 0.01$ \\
C2& 14.7 & 0.94 & 33.5 & $0.81\pm 0.01$ & $1.00\pm 0.01$ \\
C3& 16.3 & 1.18 & 37.9 & $0.79\pm 0.03$ & $0.85\pm 0.02$ \\
\hline
\hline
\end{tabular}
\end{center}
\label{param}
\end{table}

\subsection{Observations and data procesing}

The data analysed here correspond to the period September 1999-January
2001. The experiment was running continuously except for periods of
testing, maintenance and technical failures, or when the atmospheric
conditions were extreme (strong winds or rain). After careful
inspection of the data, we selected  $\sim 100$ days of good-quality for the
analyses of this paper.
Table~\ref{obs} presents a summary of the observations indicating the
number of useful maps selected in each one of the five different instrumental
settings of COSMO15 in this period.

\begin{table}
\caption{Observations analysed in this work}
\begin{center}
\begin{tabular}{ccccc}
\hline
\hline
Epoch      & Range in Decl.  & C1 & C2 & C3\\
(dd/mm/yy) &     (deg)      &   \multicolumn{3}{c}{(Number of days)} \\
\hline
27/09/99-08/11/99 & 16.7-36.7 &  3 &  4 &  5 \\
01/12/00-15/02/00 & 23.7-43.7 & 13 & 10 &  7 \\
20/02/00-15/03/00 & 27.4-47.4 & 10 &  9 &  7 \\
22/03/00-18/09/00 & 24.8-44.8 & 56 & 56 & 45 \\
06/11/00-16/01/00 & 26.4-46.4 & 36 & 27 & 20 \\
\hline
\hline
\end{tabular}
\end{center}
\label{obs}
\end{table}

The basic steps in  data processing include filtering out data affected by technical failures
or proximity of the Sun and Moon, removal of differential atmospheric emission, absolute
calibration and construction of a daily map for each of the three channels. The first steps of
data processing are done in the Fourier space as explained in Gallegos et al. The dominant
component in each scan is a spin-synchronous signal due to a combination of differential
atmospheric emission and ground pick-up.  As shown in Gallegos et al. (2001), this component
appears in the data as large scale modulations, so it can be accurately removed using a
combination of low order sinusoidal functions. In particular, for the map-making procedure of
this paper, we have subtracted $7$ harmonics. Further additional components due to changes in
the atmospheric  conditions during the day are also eliminated by a long period fit to each
point of the scans over the day. These two processes completely eliminate the response of the
experiment to temperature fluctuations on scales $\ge 5$ degrees.

\begin{figure}
\begin{center}
\includegraphics[width=9.0cm,angle=0]{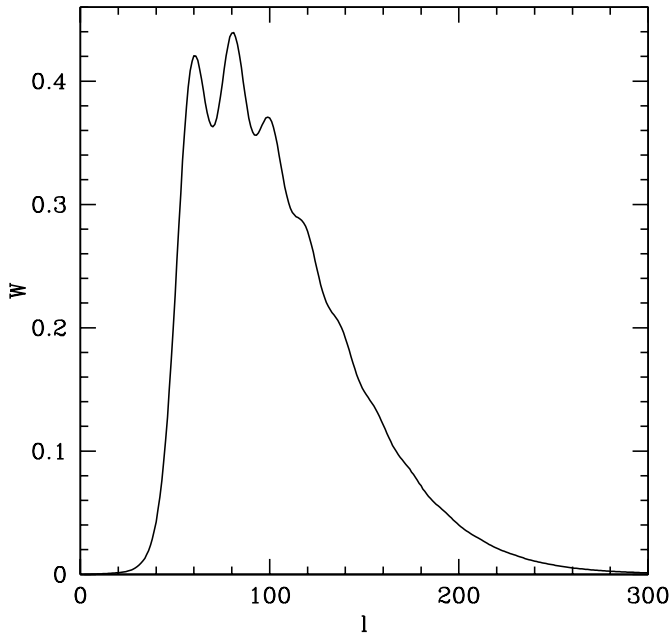}
\caption{Effective window function of the COSMOSOMAS maps.}
\label{ventana}
\end{center}
\end{figure}

We follow a numerical approach to evaluate the response of the experiment  to  different
angular scales. This is done simulating the reponse to a point source  after passing through
the scanning strategy of the experiment.   Given that the data have been obtained with five
different instrumental settings  (see Table~\ref{obs}), we have defined an effective window
function as a weighted  mean value of the  window functions  associated with each setting. For
the different window functions the integral under the function is used and for different
pointings the weight is proportional to the number of days observed. Figure~\ref{ventana}
presents this effective window function of the experiment where  we have taken into account
that the final maps are convolved to a common resolution of 1.12 degrees  (see next
sections). 

\section{Maps}

\subsection{Daily maps and astronomical calibration}

Using the strong astronomical radio sources Cyg~A (RA=299.87$^\circ$,
Dec.=40.73$^\circ$, Eq. 2000) and Tau~A (RA=83.64$^\circ$,
Dec.=22.01$^\circ$, Eq. 2000) we accurately obtain the optical configuration
which in turn is used to determine the path swept out by the circular
scanning.  With this the celestial coordinates of data in each position of
the scans can by calculated and used to project them into a grid with
$\frac{1}{3}\times \frac{1}{3}$ degrees in RA and Dec. This grid was choosen
to adequately sample the beam. A daily map with full coverage in RA and 20
degrees width in declination is obtained for each channel. 

The calibration process implies the determination of the angular response
and the conversion from counts to physical units of temperature.  A complete
calibration of the experiment is performed on a daily basis using the same
standard astronomical sources used in the pointing determination; Cyg~A and
Tau~A. Both are stable and sufficiently strong to be used as absolute
calibrators of the daily maps. Their fluxes at the COSMOSOMAS frequencies
have been computed following the model by Baars et al. (1977). This model
has been checked against the fluxes measured for these sources by
WMAP\footnote{The WMAP analysed in this paper are the first year full sky
maps (see http://lambda.gsfc.nasa.gov/product/map/dr1/\\imaps\_frequency.cfm).}
 in the K, Ka, Q and V channels,
and with the direct measurement of Cyg~A at 15 GHz by Baker et al. (1977).
We estimate the uncertainty in calibration of our data is $\le 10$ \%. The
FWHM of the beam is determined from the radial profile in Dec. and RA. of
these calibration sources. Average values of the FWHMs measured in all daily
maps and their errors are listed in Table~\ref{param}. The conversion factor
is then established from source flux and beam FWHM and apparent amplitude in
counts. In the calibrated daily maps we find typical sensitivities are
0.5-1.0 mK per beam.

\subsection{Noise analysis}

\begin{figure}
\begin{center}
\includegraphics[width=9cm,angle=0]{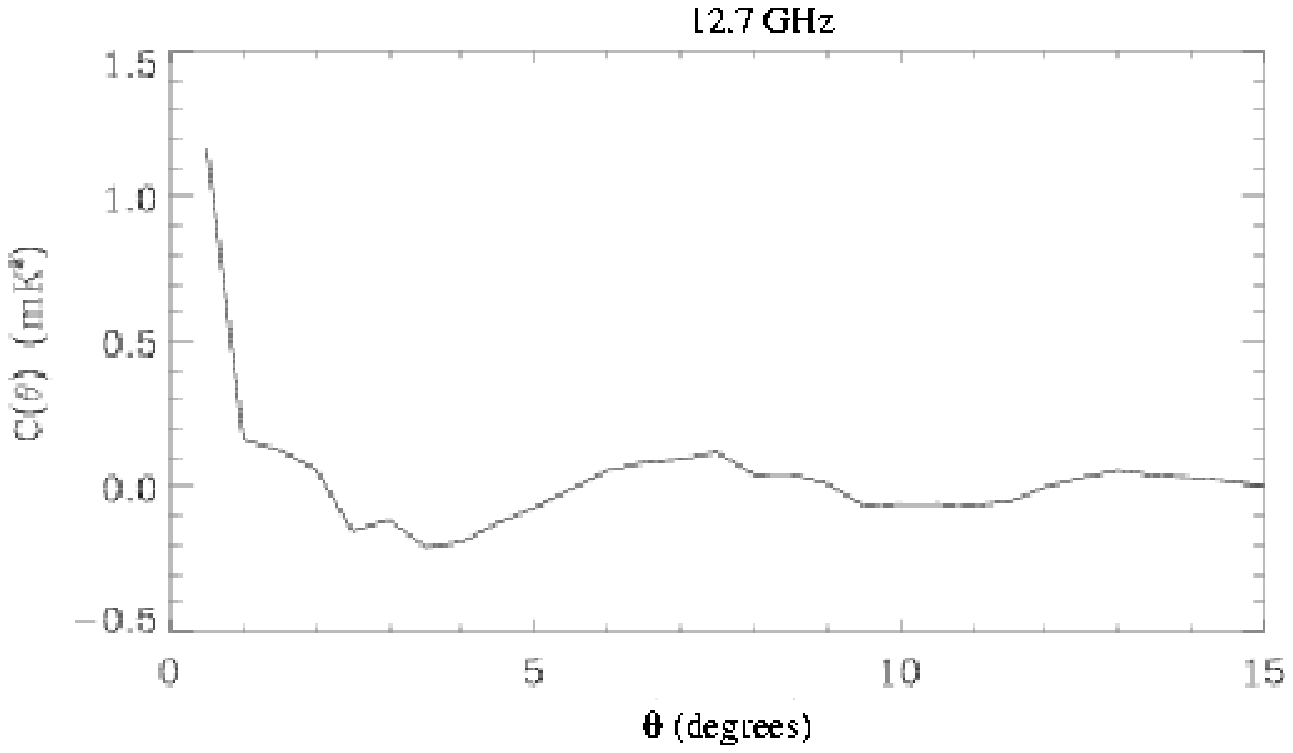}
\includegraphics[width=9cm,angle=0]{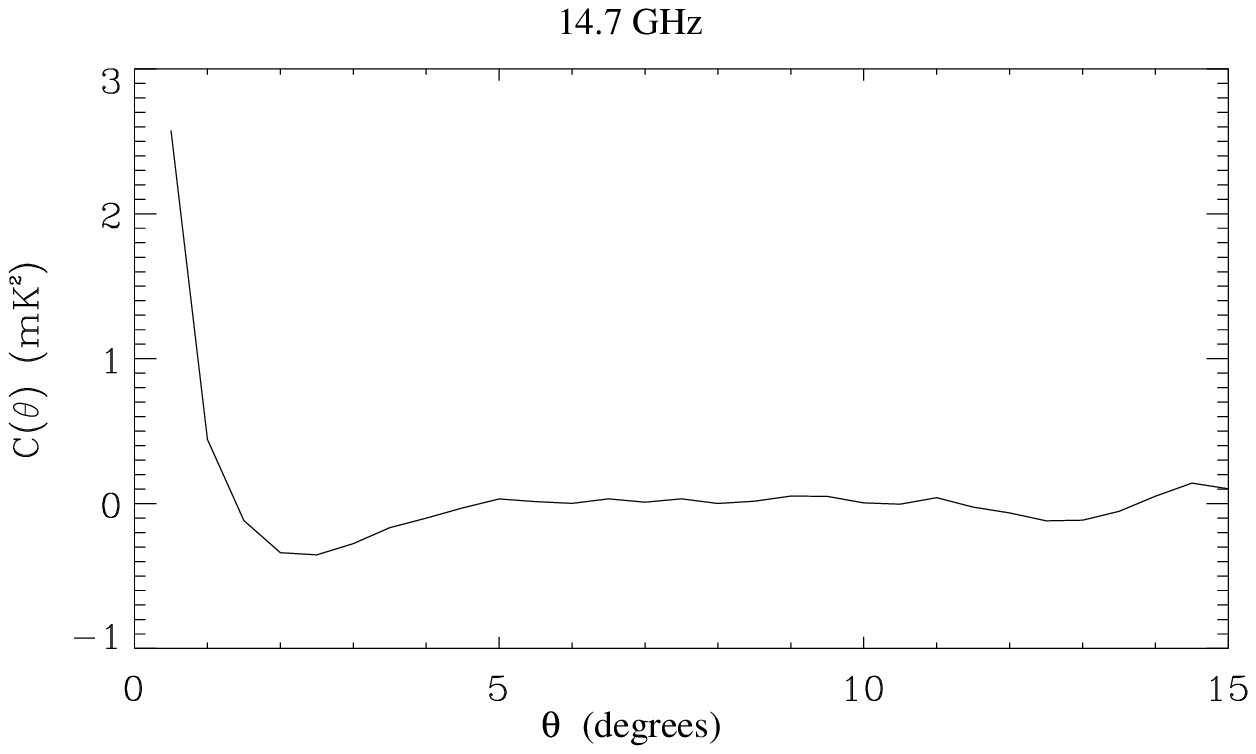}
\includegraphics[width=9cm,angle=0]{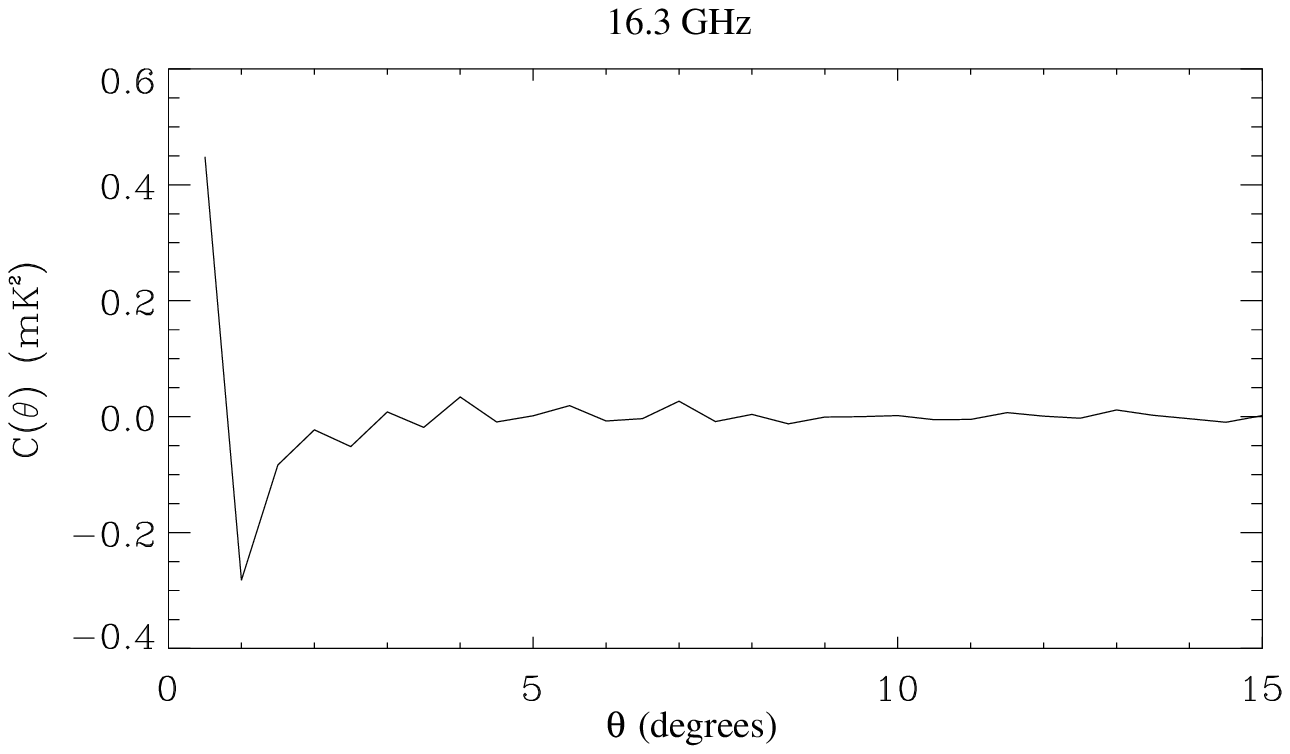}
\caption{Correlation function of one day of COSMOSOMAS data at the 
three observing frequencies.}
\label{fig_corr}
\end{center}
\end{figure}

A robust estimation of the nature and amplitude of the astronomical signals
present in the COSMOSOMAS maps require an accurate knowledge of the
statistical properties of the noise.  The noise in each daily map is
characterised by the noise covariance matrix between each pair of pixels. A
reasonable first approximation assumes that this only depends on angular
distance. This is further simplified if the noise is white, in which case
the covariance matrix is diagonal. In our case, the noise matrix can be
estimated from the analysis of data at high galactic latitude in single day
observations. In such daily maps the relative contribution of the
astronomical signals is small compared to the dominant noise component.
Figure~\ref{fig_corr} shows the correlation functions (assuming that the
covariance between any pair of pixels is only a function of the distance)
evaluated from one typical day of data in the three channels of COSMOSOMAS.
For this analysis, we selected the region enclosed in a square 
RA=(115$^\circ$, 267$^\circ$) and Dec.=(25$^\circ$, 45$^\circ$), excluding
also the pixels affected by the presence of radiosources according to the
mask described in Section~4. The presence of non diagonal elements in the
covariance matrix is revealed by non-zero values in the correlation function
at angles different from zero. Whether this is due to the instrument or to
the atmosphere is analysed below.  

Because the observations are simultaneous in the three channels, some degree
of correlated signal between them is expected due to common parts in the
receiver system and contamination by atmospheric signals. We have estimated
this by cross correlating the simultaneous observations in the three
channels.  Table~\ref{tab_covar} presents the average noise values
(diagonal) and square root of the average covariances between daily maps in
each channel over the whole period of observations. The listed
values are derived by the following expression
\[
Cov(k,k') = \frac{\sum_{i,i'}\sum_{j,j'} T_{ij}^{k}
T_{i'j'}^{k'}}{\sum_{i,i'}\sum_{j,j'} 1},  \qquad  k,k'=C1,C2,C3 \] 
where $T_{ij}^k$ corresponds to 
the temperature of the band $k$ (= C1, C2, C3), in pixel $i$ and 
in map $j$ ($j$ runs over all daily maps). This correlated
signal between channels extends up to angular scales of 5 degrees. We
attribute this daily correlation to the presence of
atmospheric noise which still persists after the filtering of the low order
harmonics.

\begin{table}
\caption{Average values for the noise and cross-correlations per pixel in the daily COSMOSOMAS 
maps.}
\begin{center}
\begin{tabular}{cccc}
\hline
\hline
& C1 & C2 & C3 \\
 & (mK) & (mK) & (mK) \\
\hline
C1& 2.11$\pm 0.73$ & $0.50\pm 0.10$ & $0.42\pm 0.13$ \\
C2& - & 2.03$\pm 0.92$ & $0.42\pm 0.11$ \\
C3& - & - & 3.30$\pm 1.12$ \\
\hline
\hline
\end{tabular}
\end{center}
\label{tab_covar}
\end{table}

We have checked for the presence of possible correlations between
consecutive daily maps and conclude that the correlation is below 1 \%. This
is at the level expected from astronomical signals. We conclude
that the atmospheric noise in the individual COSMOSOMAS maps is uncorrelated
from day to day, and thus each daily map is considered as independent in the 
stacking of the  data. 

\subsection{Stacked maps}

\begin{figure}
\begin{center}
\includegraphics[width=9.0cm,angle=0]{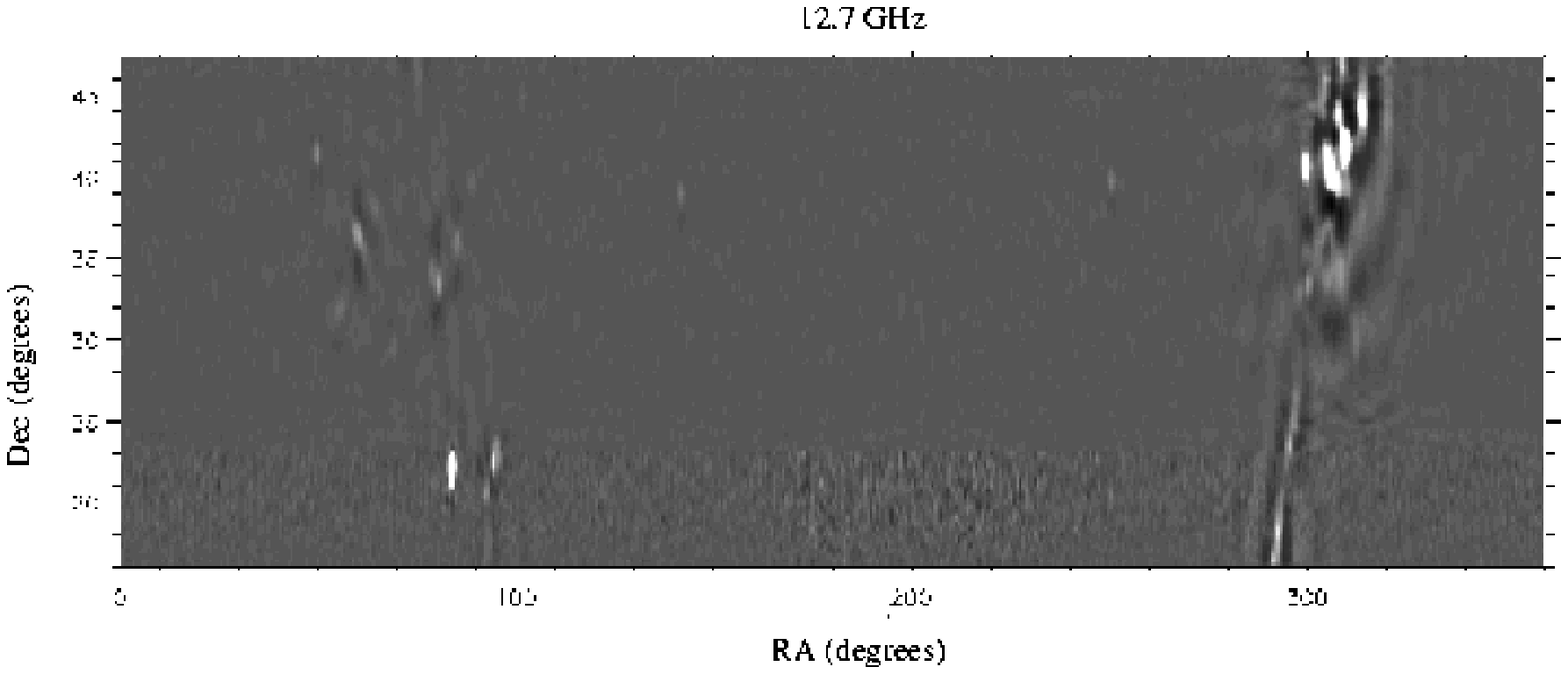}
\includegraphics[width=9.0cm,angle=0]{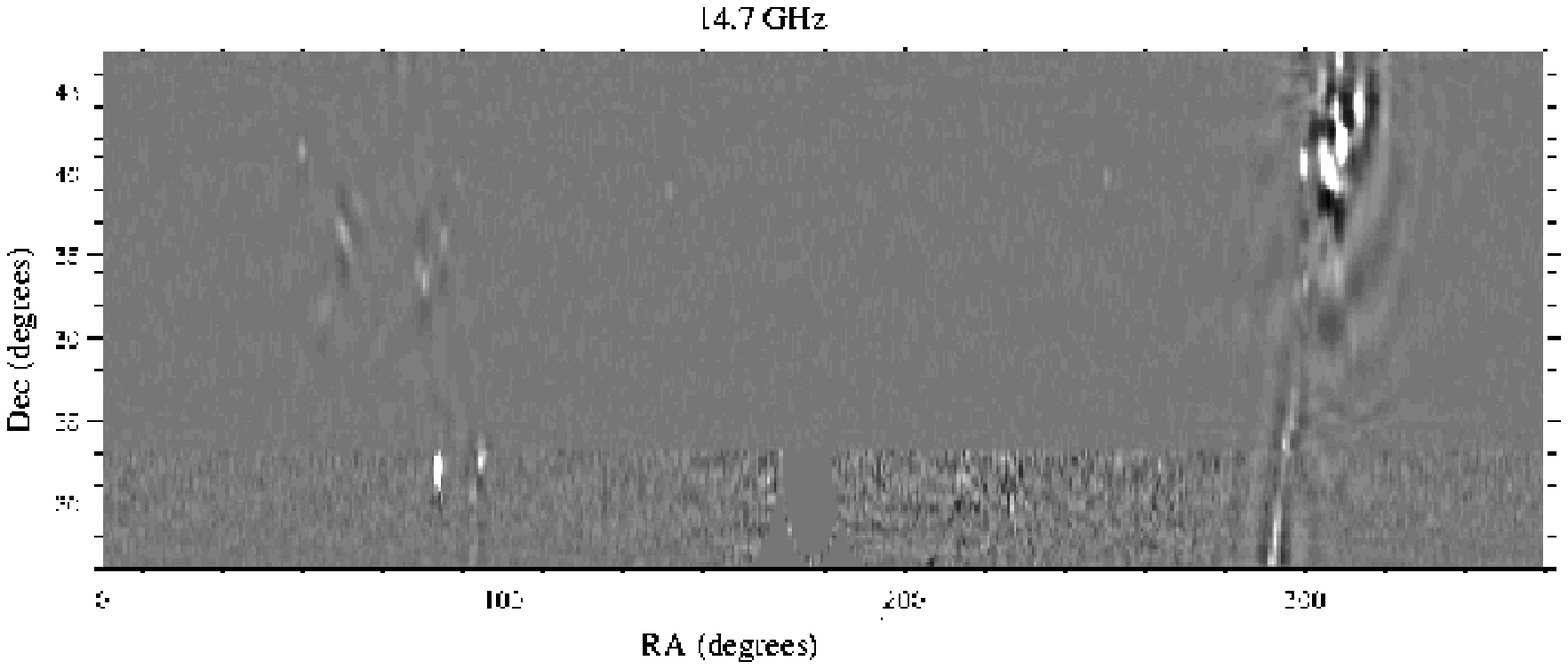}
\includegraphics[width=9.0cm,angle=0]{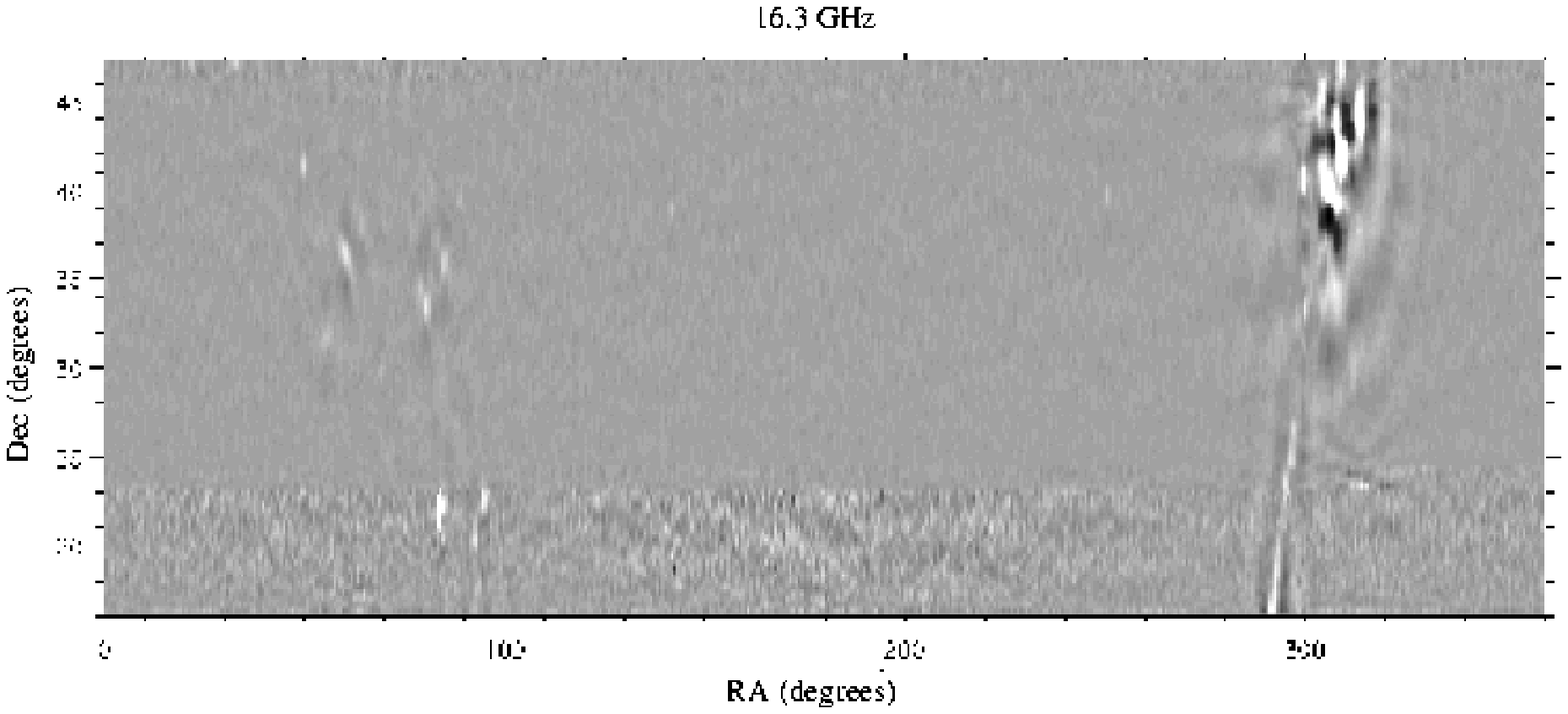}
\caption{Stacked COSMOSOMAS maps at 12.7 ($top$), 14.7 ($middle$) and 16.3
($bottom$) GHz.}
\label{maps}
\end{center}
\end{figure}

For each channel, the $n$ daily maps are stacked together to get final maps
with improved signal to noise ratio by a factor $\sqrt{n}$. These final maps
are presented in Figure~\ref{maps}. They are complete in right ascension and
cover from 16.7 to 47.4 degrees in declination. The conspicuous structures
at RA$\sim 300$ degrees correspond to the main crossing of the galactic
plane. The two strong sources Tau~A and Cyg~A are clearly visible. Other
sources are also present in the maps and will be described below. The
regions with higher sensitivity are concentrated in the band $25^\circ\le$
Dec. $\le 45^\circ$. At Dec. $\le 16.7^\circ$ the number of observations is
smaller and the noise is higher as is easily appreciated in
Figure~\ref{maps}.  The errors are calculated from the dispersion of all the
measurements contributing to a given pixel of the map. The mean noise in a
wide region (120$^\circ\le$ RA $\le 270^\circ$, 25$^\circ\le $ Dec. $\le
45^\circ$) excluding the galactic plane, are in 173, 185 and 329
$\mu$K per pixel for channels C1, C2 and C3 respectively; these are
translated into 49, 59 and 115 $\mu$K beam$^{-1}$. The rms of the three maps
computed in the same region are 90, 90 and 116 $\mu$K beam$^{-1}$. These
larger values of the rms as compared with the noise indicates the presence
at high galactic latitudes of clear statistical signals in each map. The
evaluation and origin of these signals is considered in the following
sections.  

\begin{figure}
\begin{center}
\includegraphics[width=5.5cm,angle=0]{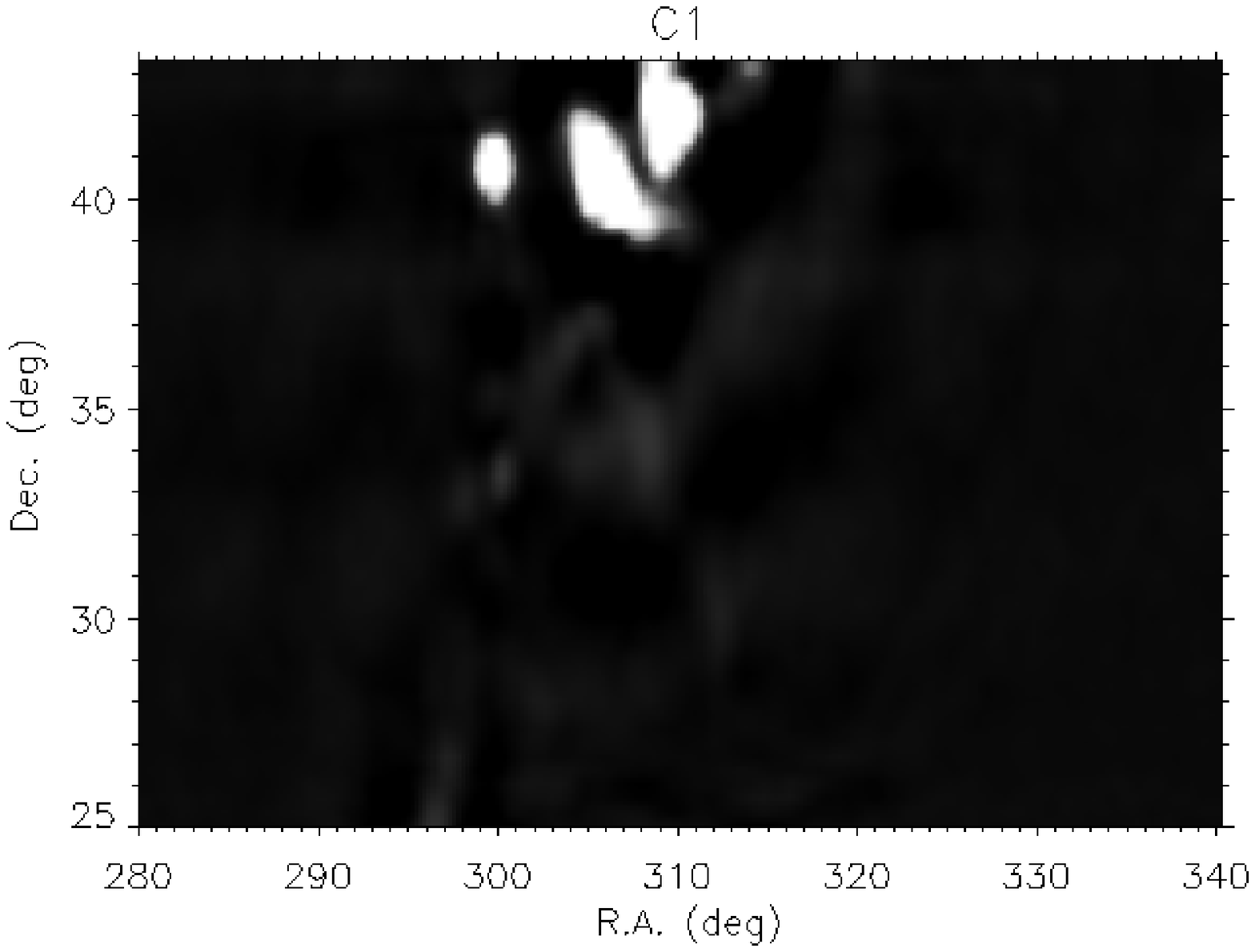}
\includegraphics[width=5.5cm,angle=0]{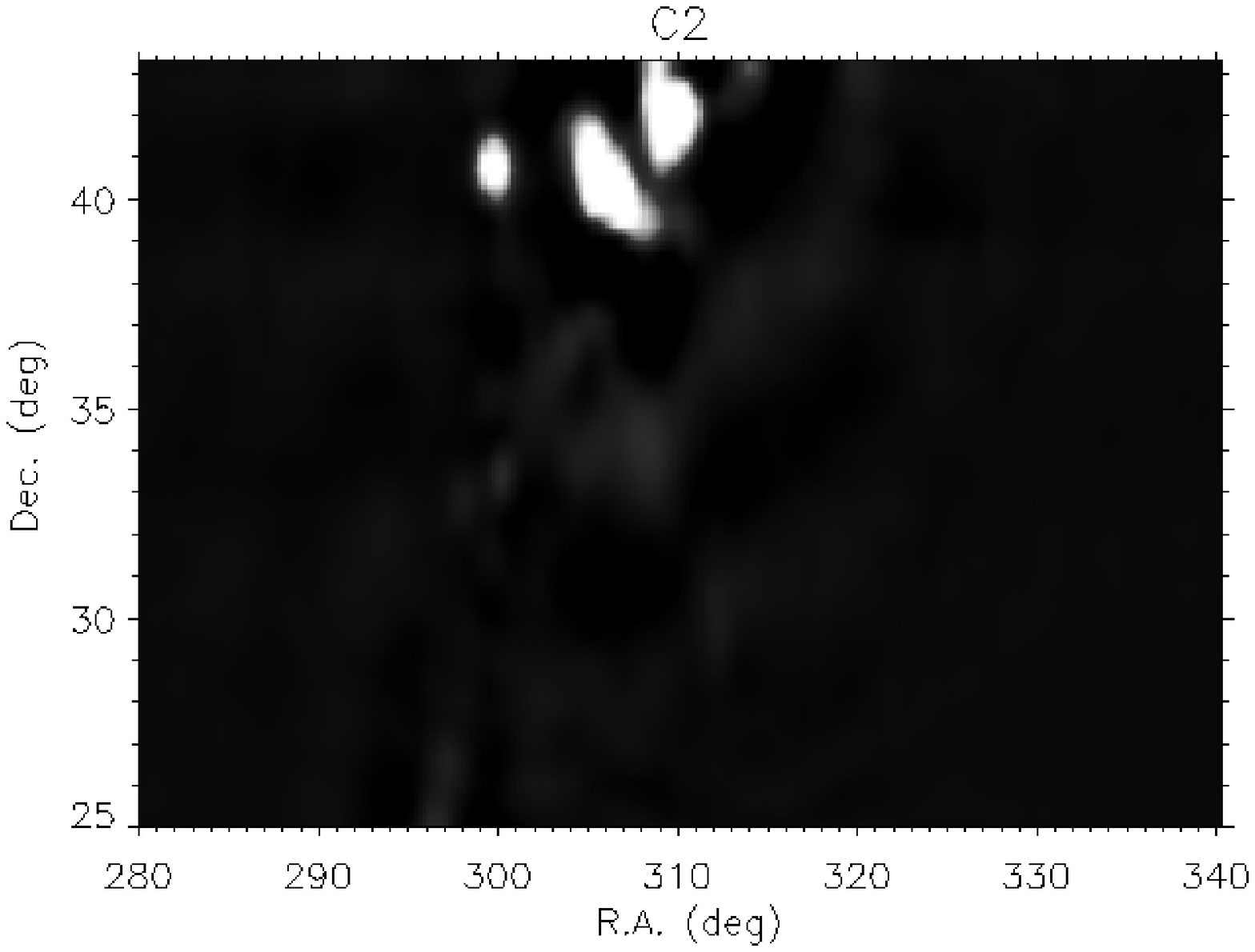}
\includegraphics[width=5.5cm,angle=0]{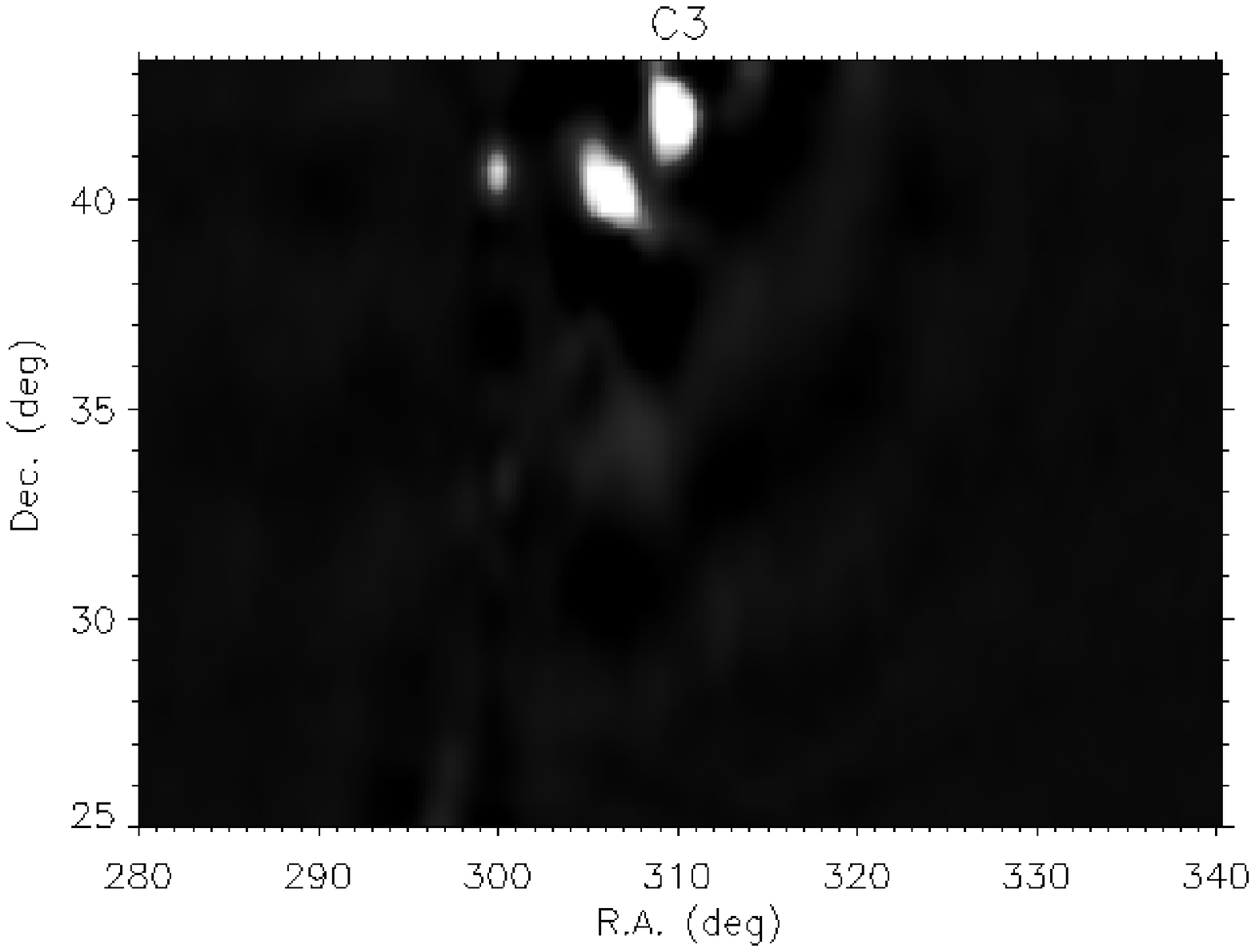}
\includegraphics[width=5.5cm,angle=0]{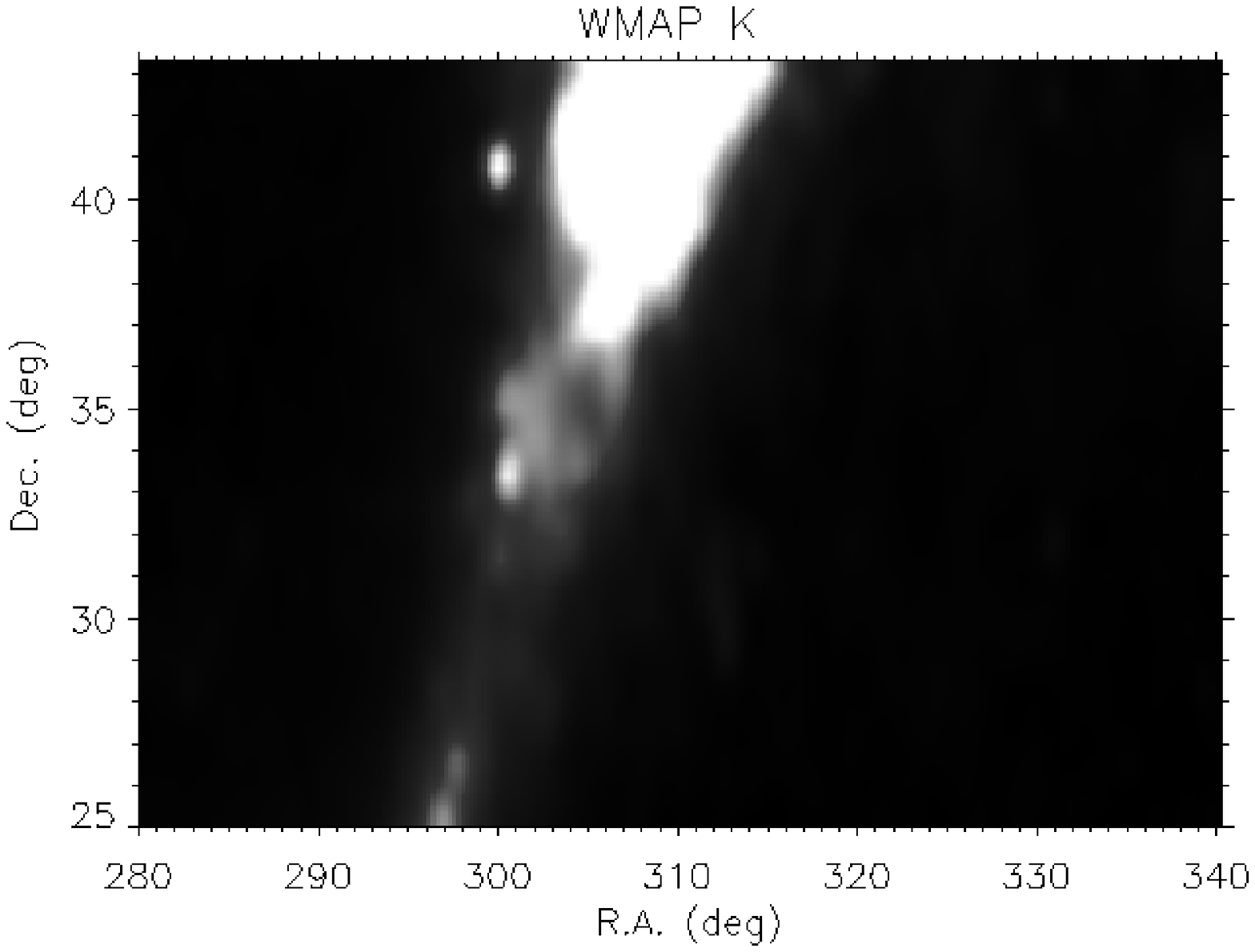}
\includegraphics[width=5.5cm,angle=0]{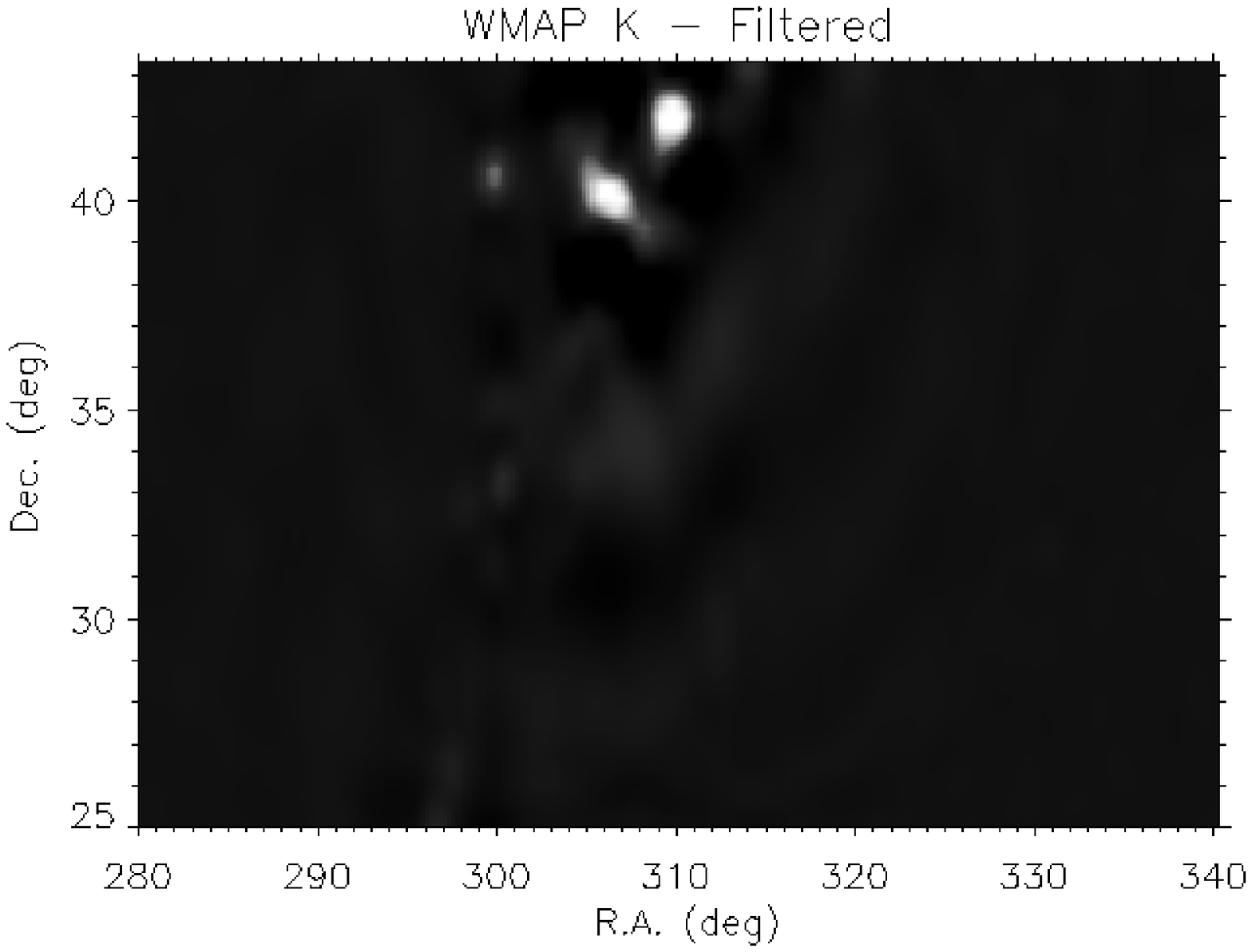}
\caption{From top to bottom: the Galactic plane region observed by the three COSMOSOMAS
channels, and the lowest frequency channel of WMAP (before and after
filtering according to the observing strategy of  COSMOSOMAS).}
\label{fig_wmap_vs_cosmo_galactic_plane}
\end{center}
\end{figure}

\begin{figure}
\begin{center}
\includegraphics[width=5.cm,angle=0]{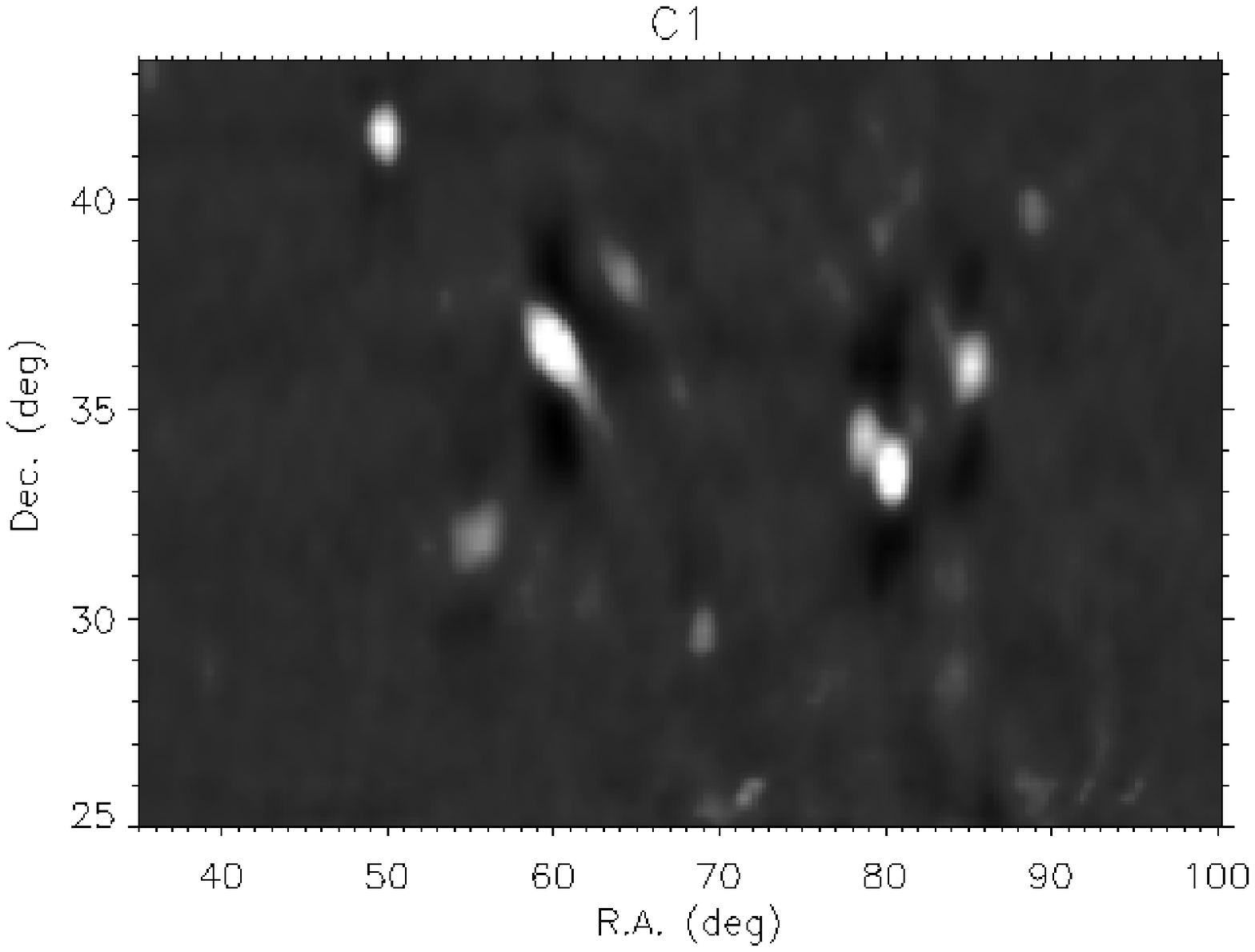}
\includegraphics[width=5.cm,angle=0]{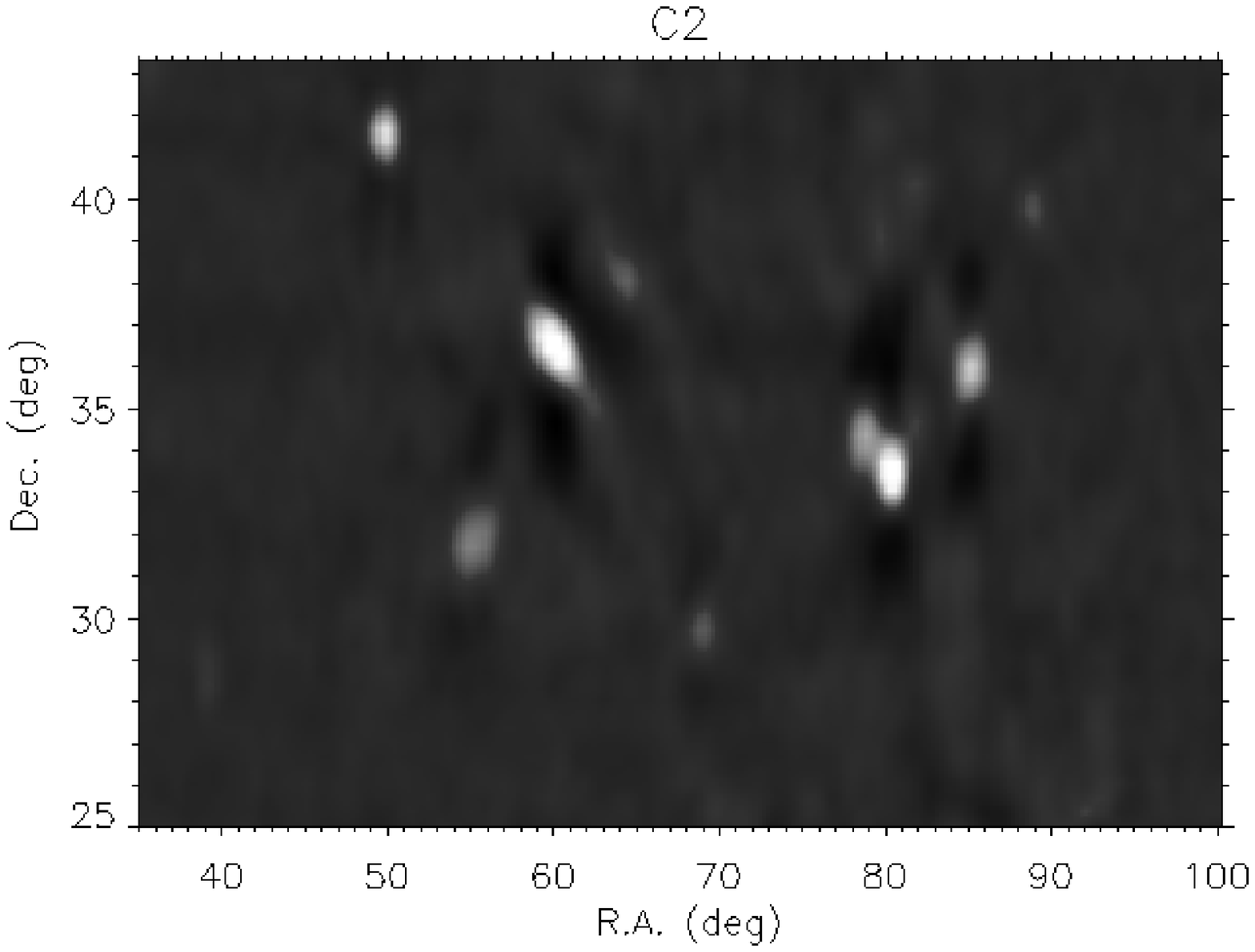}
\includegraphics[width=5.cm,angle=0]{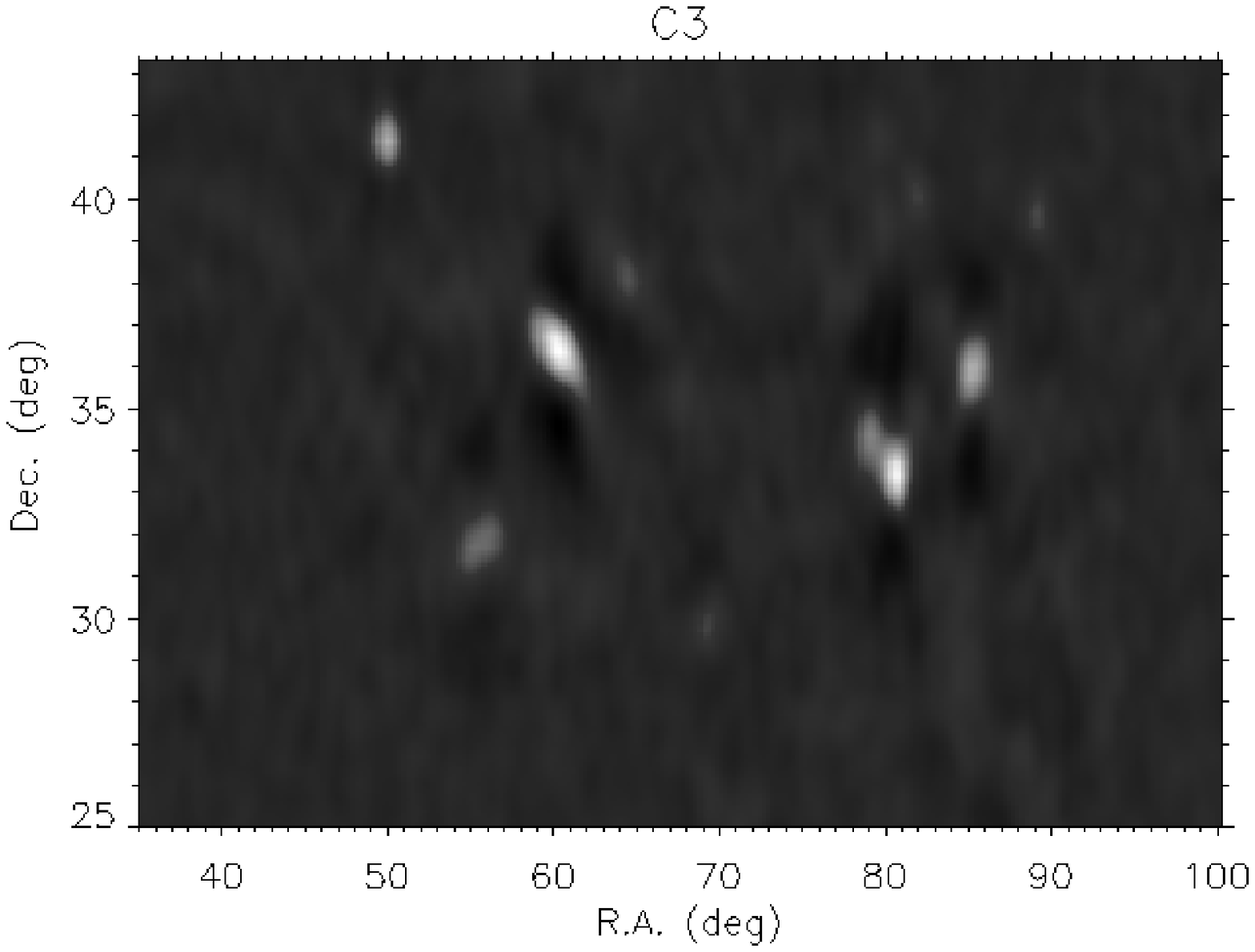}
\includegraphics[width=5.cm,angle=0]{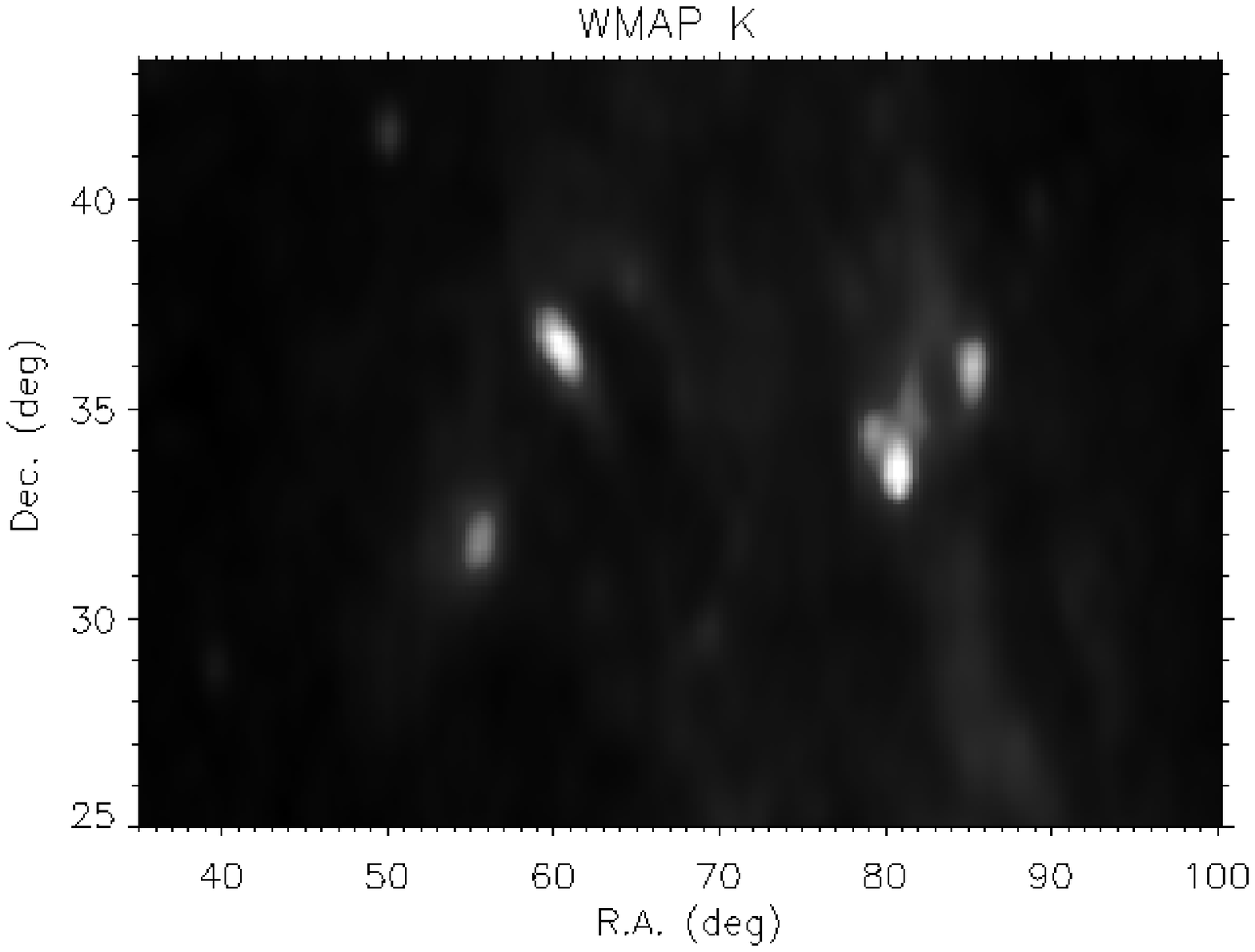}
\includegraphics[width=5.cm,angle=0]{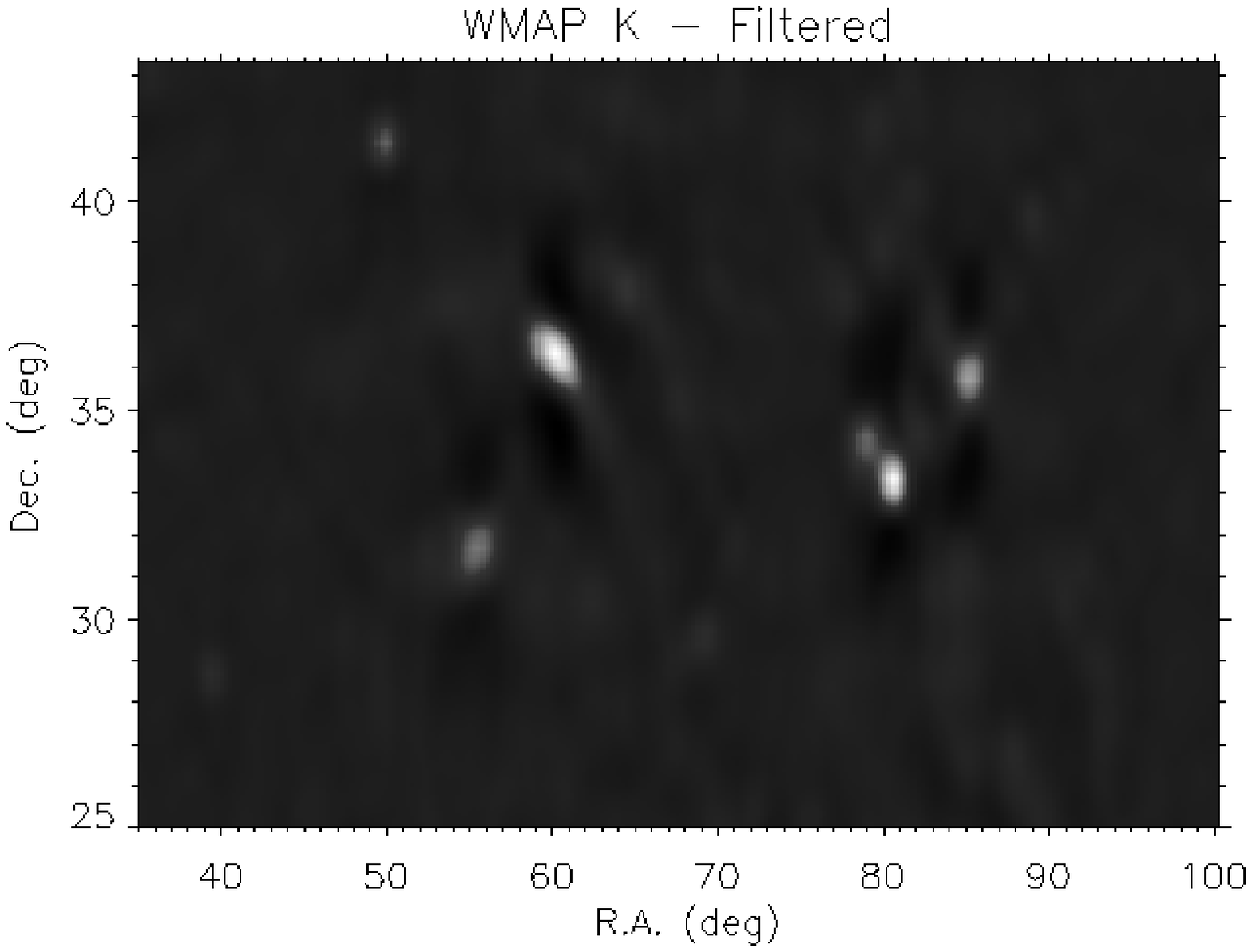}
\caption{As in the previous figure, but for a field towards 
the Auriga-Perseus region.}
\label{fig_wmap_vs_cosmo_anticentro}
\end{center}
\end{figure}

Figures~\ref{fig_wmap_vs_cosmo_galactic_plane} and~\ref{fig_wmap_vs_cosmo_anticentro} show a
visual comparison between the COSMOSOMAS maps at the three observing frequencies, and the WMAP
map at 23 GHz (all convolved to a common circular  resolution of 1.12$^\circ$, which
corresponds to the worse resolution of those presented in Table~1). The two regions displayed
are located in the galactic plane, and the prominent structures are mainly due to diffuse
emission from our galaxy. The remarkable agreement between the structures observed by
COSMOSOMAS and WMAP supports the reliability of the former. Table~\ref{tab_fuentes} lists the
observed fluxes of some of  the strongest radio sources identified in the COSMOSOMAS maps. The
fluxes are determined by the same method used in Watson et al (2005) where the best fitting
amplitude for the expected modelled source profile to the data is found. For a comparison
Table~\ref{tab_fuentes} presents also the fluxes measured for these sources in the two low
frequency WMAP maps (Bennett et al. 2003). COSMOSOMAS and WMAP data are not coeval, and so
this comparison is only indicative for variable sources (3C84, 4C39.25, 3C345, etc). As
expected in the frequency range covered by COSMOSOMAS and WMAP most of the sources have a flat
spectrum. The contributions of unresolved sources will be quantified in Section~4.

\begin{table*}
\caption[Point sources]{Fluxes of several point sources identified in
the COSMOSOMAS C1, C2 and C3 maps. WMAP fluxes in the K and Ka
channels (taken from Bennett et al. 2003) are also listed. Errors and upper limits are 68 \%  and 95 \% c.l. respectively.}
\begin{center}
\begin{tabular}{rrrrrrr}
\hline
& &  \multicolumn{3}{c}{COSMOSOMAS} &  \multicolumn{2}{c}{WMAP} \\
ID& RA, Dec  & F$_{C1}$  & F$_{C2}$  & F$_{C3}$     &  
	F$_K$  &  F$_{Ka}$   \\
	& (J2000) & (Jy) & (Jy) & (Jy) & (Jy) & (Jy) \\
\hline
0234+28  &(39.5$^\circ$, 28.8$^\circ$) & $ 2.6\pm0.3   $ & $ 4.1\pm0.5   $ &
$ 6.0\pm1.0     $ & $4.6\pm0.5	$ & $ 5.0\pm0.6  $ \\
0316+41 (3C84)     &(50.0$^\circ$, 41.5$^\circ$) & $ 17.4\pm1.8  $ & $
19.6\pm2.0  $ & $ 16.8\pm1.9   $ & $12.3\pm1.2  $ & $  8.3\pm0.9   $ \\
0433+29  &(69.3$^\circ$, 29.7$^\circ$) & $ 7.7\pm0.8   $ & $ 8.9\pm1.0   $ &
$ 9.8\pm1.0     $ & $5.3\pm0.5	$ & $ 4.3\pm0.5   $ \\
0738+31  &(115.3$^\circ$, 31.2$^\circ$)& $ 2.4\pm0.4   $ & $ 1.0\pm0.4   
$ & $< 1.4$  & $1.6\pm0.2	$ & $ 1.0\pm0.3   $ \\
0923+39 (4C39.25) &(141.8$^\circ$, 39.0$^\circ$)& $ 9.1\pm1.0   $ &
$9.5\pm1.0  $ & $7.0\pm1.2     $ & $7.8\pm0.8	$ & $ 6.4\pm0.7   $ \\
1328+30 (3C286) &(202.8$^\circ$, 30.5$^\circ$)& $ 2.4\pm0.3   $ & $ 3.0\pm0.4   
$ & $ 2.7\pm0.9     $ & $1.5\pm0.1	$ & $ 0.7\pm0.2 $ \\
1640+33* & (250.0$^\circ$, 33.5$^\circ$) &  $1.2\pm0.3 $ & $  1.6\pm0.4   $ & $ 1.9\pm0.9   $ & $ 1.6\pm0.2$ &  $3.1\pm0.3  $ \\
1641+39 (3C345) &  (250.7$^\circ$, 39.8$^\circ$)  & $13.1\pm1.3 $ & $  14.0\pm1.4  $ & $ 12.8\pm1.5  $ & $  8.2\pm0.8$ &  $  6.9\pm0.7 $ \\
1611+343* &(243.4$^\circ$, 34.2$^\circ$)  & $4.4\pm0.5 $ & $   4.5\pm0.6   $ & $ 4.2\pm0.9   $ & $ 4.5\pm0.5$ &  $  4.7\pm0.5 $ \\
\hline
\medskip
\end{tabular}
\label{tab_fuentes}
\end{center}
\end{table*}

\section{Cross correlation analysis}

The limited frequency range spanned by the three COSMOSOMAS channels
analysed here, do not allow an accurate direct separation of the various
astronomical components present in the data, namely CMB and the different
galactic foregrounds. To evaluate the presence of each astronomical signal
we have applied cross-correlation techniques between the COSMOSOMAS maps and
existing templates considered as good tracers of the different galactic
foregrounds.  These
cross-correlation techniques allow an estimation of which fraction of the
total signal contained in a given map is correlated with a reference
template. The main assumption is that the spatial structure in the
reference and the map under analysis is the same apart from a scaling factor
in amplitude. 

\subsection{Templates}

The maps used as templates are the 408 MHz (Haslam et al.
1982), 1420 MHz (Reich \& Reich 1986),
DIRBE\footnote{http://lambda.gsfc.nasa.gov/product/cobe/}, 
and H$\alpha$ (Finkbeiner 2003). 
At 408 MHz the dominant diffuse foreground is synchrotron, however
the Haslam et al. map at this frequency is severely contaminated by the
contribution of point sources, so as a best tracer of the synchrotron
emission we have used the  'cleaned map' (desourced) provided on the NASA LAMBDA 
site\footnote{http://lambda.gsfc.nasa.gov/product/foreground/haslam\_408.cfm}. 
The main
properties of these templates are summarized in Table~\ref{tab_mapas_ref}.
The 1420 MHz has a contribution of both synchrotron
and free-free emission. H$\alpha$ emission is a tracer of the warm
ionized medium which also radiates via free-free emission, although an
accurate estimation of this foreground requires a correction for the
extinction by dust and the electron temperature of the medium
(Dickinson et al 2004). However, as we are interested in correlated signals at high galactic
latitudes where the dust extinction is negligible, we have not attempted
this correction. 
We only use the DIRBE maps at 100 and 240 $\mu$m as good tracers of galactic dust.
We include in the cross-correlation analysis also the first year WMAP maps,
which are dominated by CMB signal, although depending on the frequency
considered different degrees of galactic contamination exists. 

Before the correlation analyses, the WMAP and galactic
template maps have been filtered 
to the common COSMOSOMAS resolution ($1.12^\circ$), and
have been convolved with the simulated circular lock-in 
analysis and pixelated in the same $1/3^\circ$ grid used before. 
In the construction of these
templates small differences will arise with respect to the COSMOSOMAS maps
due the 3-sigma clipping algorithm implemented to reduce the effect of very
strong radio sources in the observed scans. These differences will be larger
as we get closer to the galactic plane so we will consider here
only correlation results at high galactic latitudes $|b|>30^\circ$. 
Due to the comparatively high amplitude of the galactic plane, we do not
discard the possibility of some residual systematic effects in the band 
$30^\circ < |b| < 40^\circ$.

\subsection{Method}

The cross-correlation between two maps is derived as follows.
Let $\bmath{d}$ be the data vector containing
$N$ data points. We want to fit simultaneously $n_t$ ``template maps''.
Let $\bmath{t}_i$ be the templates ($i=1,...,n_t$), and
$\bmath{\alpha}$ a vector with $n_t$ elements containing
the cross-correlation variables in which we are interested.
Then, the $\bmath{\alpha}$ vector can be derived by minimizing

\begin{equation}
\chi2 = ( \bmath{d} - \bmath{\alpha} \cdot \bmath{\hat{M}})^T
\cdot \bmath{\hat{C}}^{-1} \cdot
( \bmath{d} - \bmath{\alpha} \cdot \bmath{\hat{M}})
\end{equation}
where $\bmath{\hat{C}}$ is the covariance matrix including both the
signal and the noise, and
$\bmath{\hat{M}}$ is a $n_t \times N $ matrix built from the
template maps, and whose elements
are defined as $M_{i,j} = ( \bmath{t}_i )_j$, where $i=1,...,n_t$ and
$j=1,...,N$.
Solving for $\bmath{\alpha}$ we obtain:

\begin{equation}
\bmath{\alpha} = ( \bmath{d} \bmath{\hat{C}}^{-1} \bmath{\hat{M}}^T)
( \bmath{\hat{M}} \bmath{\hat{C}}^{-1} \bmath{\hat{M}}^T )^{-1}
\end{equation}

and the corresponding covariance matrix of the $\bmath{\alpha}$
parameters is

\begin{equation}
Cov(\bmath{\alpha}) =
( \bmath{\hat{M}} \bmath{\hat{C}}^{-1} \bmath{\hat{M}}^T )^{-1}
\end{equation}

\noindent
The variance for each parameter $\alpha_j$ is given by
the square-root of the corresponding element in the diagonal of the
covariance matrix.
Gorski et al. (1996) described this same method but in harmonic space.
Note that for the case $n_t=1$ we recover the standard equations
presented in, e.g. de Oliveira-Costa et al. (1999), for the case of a single
template.

We have conducted several tests to study the correlated component of the
noise and the CMB in the covariance matrix. Our conclusion is that both
terms are small as compared with the white component of the noise and with
the dominant foreground of each template, and so they can be ignored except
when WMAP maps are considered as templates, in which case it is necessary to
take into account the CMB component. The analysis
has been performed in declination strip 25$^\circ$-44.7$^\circ$ for
galactic latitude cuts $|b|>$ 30, 40 and 50 degrees to study the dependence of
the different contributors with galactic latitude.  

We note that the harmonic supression procedure followed in our processing of
the COSMOSOMAS data removed a large proportion of the correlations induced
by the large scale structure of our galaxy (e.g. cosec law distribution). 

\begin{table}
\caption{Galactic templates The third column indicates the dominant components
in each template (1. CMB; 2. Synchrotron; 3. Free-free; 4. Dust).}
\label{tab_mapas_ref}
{\tiny 
\begin{center}
\begin{tabular}{lccc}
\hline
Map& Frequency & Component & Sky coverage \\
\hline
\hline
Haslam et al. (1982)  & 408 MHz & 2 &Full sky\\
Reich \& Reich (1986) &1420 MHz & 2, 3&$-19^\circ \leq \delta $ $
\leq +90^\circ $\\
DIRBE& 100, 240 $ \mu$m & 4 &Full sky\\
Finkbeiner (2003) & H$_{\alpha}$& 3&Full sky \\
WMAP & $23, 33, 41, 61, 94$~GHz & 1, 2, 3, 4 &Full sky \\
\hline
\end{tabular}
\end{center}
}
\end{table}

\begin{figure}
\begin{center}
\includegraphics[width=8.0cm,angle=0]{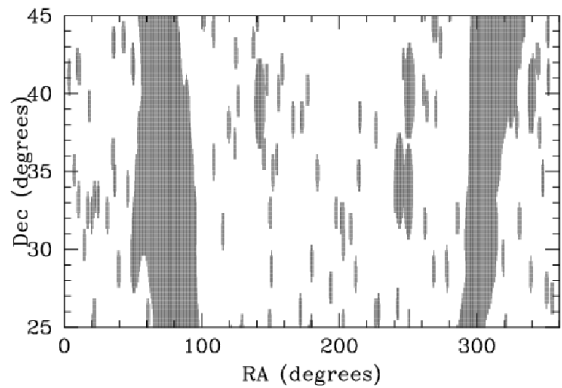}
\caption{Mask of discrete sources and galactic emission used in the correlation
analysis. }
\label{fig_mascaras}
\end{center}
\end{figure}

We assume a model in which each COSMOSOMAS and WMAP map contains a
superposition of CMB, diffuse galactic fluctuations, and noise.  Due
to the high sensitivity of the WMAP maps, their structures are
dominated largely by astronomical signal.  The WMAP collaboration
provides several masks which account for sky regions near the Galactic
plane or intense radio-sources. We have used the so-called Kp0 mask
(Bennett et al. 2003) which eliminates 21 \% of the area of the sky,
mostly situated in the galactic plane. This mask excludes also a
region within  0.6$^\circ$ around the $\sim 700$ point
sources contained in the catalogue by Stickel et al. (1994), and in
catalogues of AGNs and X-ray sources (Bennett et al. 2003). We have also masked
an additional
source located at RA=250.33$^\circ$, Dec.=33.66$^\circ$ (see Table~4) which has $\sim 1.4$
and $\sim 0.5$ Jy in the C1  and in WMAP\_K maps respectively.  To account
for the large beam of COSMOSOMAS as compared with WMAP, we have
extended the Kp0 mask by 1 pixel ($0.33^{\circ}\times 0.33^{\circ}$)
surrounding all of the masked sources. For the three most intense sources
(3C45, 4C39.25 and DA406) the mask was extended to cover a radius of
$3$~degres, in order to remove scanning artifacts larger than the noise.
Figure~\ref{fig_mascaras} presents this mask. 

\begin{table}
\caption{Correlated signal between the WMAP and COSMOSOMAS maps for galactic
latitude cuts (from top to bottom) of $|b|>30^\circ$, $>40^\circ$ and 
$>50^\circ$.  Numbers correspond to the product of the 
coupling constant $\alpha$ and the rms of the template WMAP map 
$\sigma_{WMAP}$. All values are in $\mu$K.}
{\tiny
\begin{center}
\begin{tabular}{lcccc}
\hline
 \multicolumn{2}{c}{WMAP}& \multicolumn{3}{c}{COSMOSOMAS}\\
\\
Channel& $\sigma_{WMAP}$ &  C1& C2 & C3 \\
\hline
 \multicolumn{5}{c}{$|b|>30^\circ$\hspace{1cm} }\\
\hline
WMAP\_K & 33.6 & $35.4\pm 1.1$ & $34.3\pm 1.2$ & $38.3\pm 2.5$ \\
WMAP\_Ka& 30.1 & $31.5\pm 1.1$ & $28.4\pm 1.2$ & $31.9\pm 2.5$ \\
WMAP\_Q & 30.0 & $29.7\pm 1.2$ & $27.6\pm 1.2$ & $31.0\pm 2.5$ \\
WMAP\_V & 29.0 & $27.3\pm 1.1$ & $25.9\pm 1.2$ & $26.3\pm 2.5$ \\
WMAP\_W & 28.6 & $26.3\pm 1.1$ & $25.3\pm 1.2$ & $27.4\pm 2.5$ \\

\hline
 \multicolumn{5}{c}{$|b|>40^\circ$\hspace{1cm} }\\
\hline
WMAP\_K & 33.5 & $38.6\pm 1.4$ & $36.3\pm 1.5$ & $42.5\pm 2.9$ \\
WMAP\_Ka& 30.0 & $34.1\pm 1.4$ & $30.0\pm 1.5$ & $35.9\pm 2.9$ \\
WMAP\_Q & 30.0 & $32.4\pm 1.4$ & $30.0\pm 1.5$ & $34.8\pm 2.9$ \\
WMAP\_V & 29.0 & $30.7\pm 1.4$ & $28.2\pm 1.5$ & $30.2\pm 2.9$ \\
WMAP\_W & 28.7 & $29.8\pm 1.4$ & $27.9\pm 1.5$ & $32.4\pm 2.9$ \\

\hline
 \multicolumn{5}{c}{$|b|>50^\circ$\hspace{1cm}  }\\
\hline
WMAP\_K & 33.6 & $38.0\pm 1.5$ & $33.8\pm 1.6$ & $42.3\pm 3.2$ \\
WMAP\_Ka& 30.3 & $35.3\pm 1.5$ & $28.4\pm 1.6$ & $36.0\pm 3.2$ \\
WMAP\_Q & 30.6 & $33.1\pm 1.5$ & $28.8\pm 1.6$ & $36.3\pm 3.2$ \\
WMAP\_V & 29.5 & $31.4\pm 1.5$ & $26.8\pm 1.6$ & $31.1\pm 3.2$ \\
WMAP\_W & 29.0 & $31.5\pm 1.5$ & $28.0\pm 1.6$ & $33.7\pm 3.2$ \\

\hline
\end{tabular}
\label{tab_senal_wmap}
\end{center}
}
\end{table}

\begin{figure}
\begin{center}
\includegraphics[width=9.4cm,angle=0]{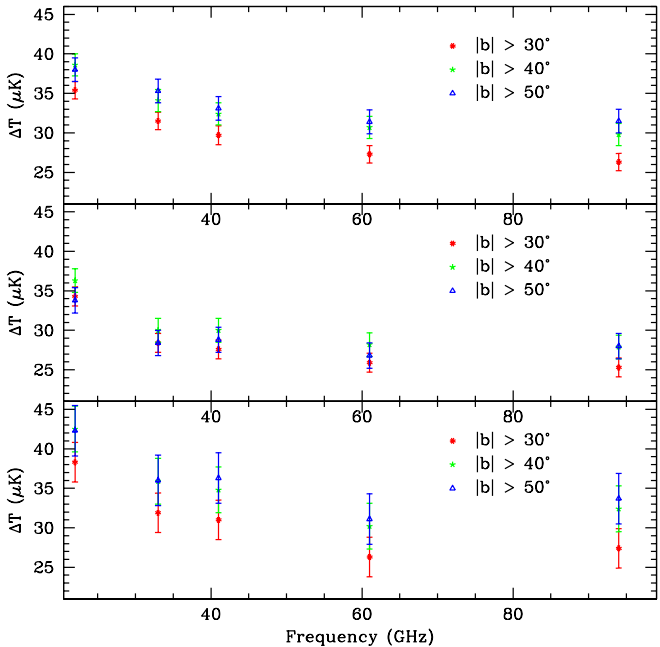}
\caption{Common signal between COSMOSOMAS ($top:$ 12.7, $middle:$
14.7, $bottom:$ 16.3 GHz) and WMAP for three galactic cuts.}
\label{fig_correl_cosmo}
\end{center}
\end{figure}

\subsection{Results}

We have obtained  all the correlations by fitting
a single template, except for the case
of cross-correlations with infrared and H$\alpha$ maps where
both maps have been fitted simultaneously since  
we expect a significant coupling between them. Indeed, 
for the case of $|b| > 30^\circ$, we find a coupling of 18 \% for
H$\alpha$-Dirbe$100$~$\mu m$, and 8 \% for
H$\alpha$-Dirbe$240$~$\mu m$.

\begin{figure}
\begin{center}
\includegraphics[width=9.4cm,angle=0]{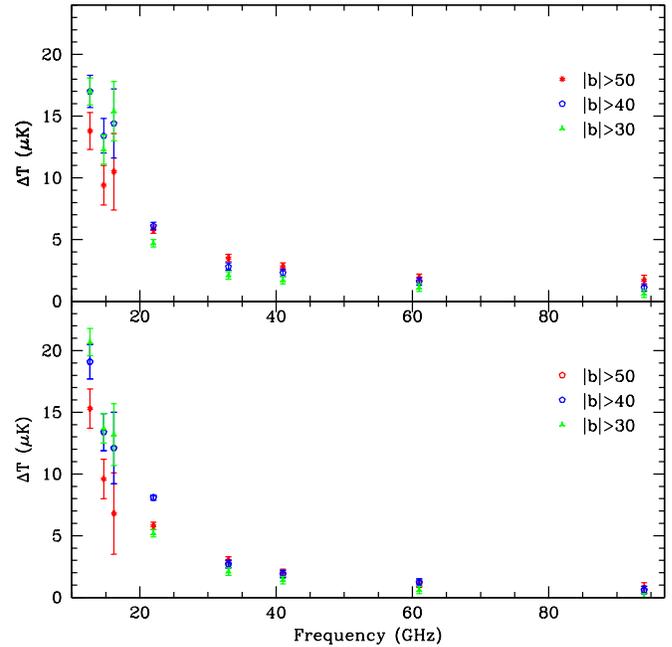}
\caption{COSMOSOMAS and WMAP correlated with the synchrotron maps at 408 ({\it top}) 
and 1420 MHz ({\it bottom}) for three 
galactic cuts.}
\label{fig_has_rei}
\end{center}
\end{figure}

\begin{figure}
\begin{center}
\includegraphics[width=9.4cm,angle=0]{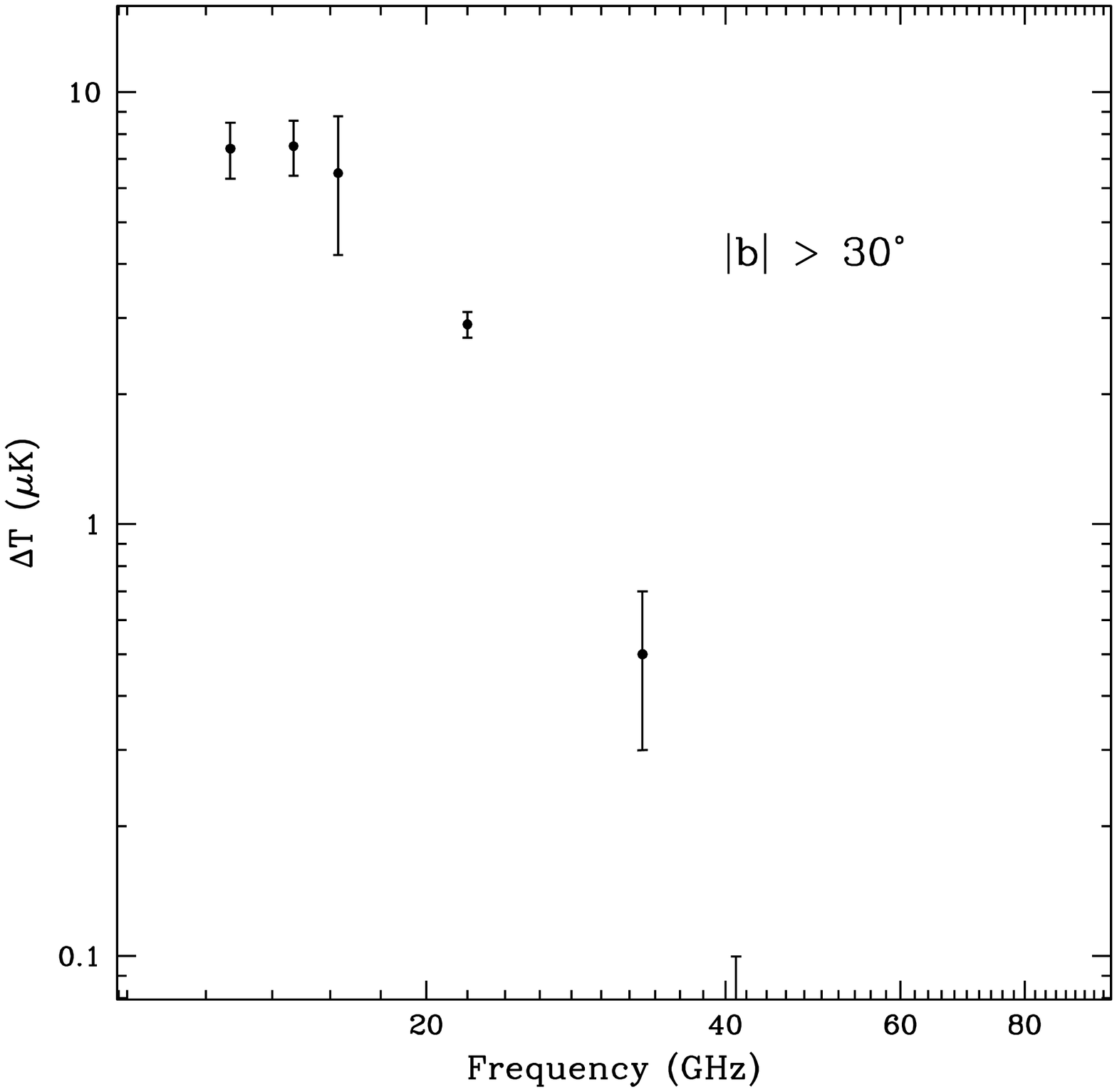}
\caption{Amplitude of the  correlated signal between COSMOSOMAS and WMAP  with
the DIRBE map at  100  $\mu$m for a galactic cut of $|b|>30^\circ$.}
\label{fig_correl_dirbe}
\end{center}
\end{figure}

The cross-correlation results between WMAP and COSMOSOMAS are shown in
Figure~\ref{fig_correl_cosmo} and Table~\ref{tab_senal_wmap}, while the
results of the correlations with the $408$~MHz, $1420$~MHz and H$\alpha$
templates are summarized in
Table~\ref{tab_templates} and Figure~\ref{fig_has_rei}.  The results of
 the cross-correlation analysis with the dust templates ($100$~$\mu m$ and 
$240$~$\mu m$) are summarized in Table~\ref{tab_dirbe}.
There are significant correlations between the
three channels of COSMOSOMAS and the two DIRBE frequencies.
Figure~\ref{fig_correl_dirbe} shows the signal correlated with the map
at 100 $\mu$m.  
The results quoted in the previous tables and figures include only the
uncertainty due to statistical errors. As a realistic test on the
significance of these errors, we performed correlations of the different
templates with randomly ordered pixel values and maps rotated with respect
the Galactic axis. While the random pixel ordenations basically confirm the
statistical errors, the dispersion of the correlations of the rotated maps
were typically 2 times larger. Thus, we will
consider as  significant those correlations with amplitude at
least 2 times the statistical error.

\begin{table*}
\caption{Correlated signal between the COSMOSOMAS - WMAP maps and three 
galactic templates. All correlation values have units of $\mu K$, and
were derived as the product of the coupling constant $\alpha$ times the
rms of the corresponding template ($\sigma_{Template}$). 
Values for 408~MHz and 1420~MHz maps were obtained using a single-template
fitting, while the values for H$\alpha$ were obtained 
from a simultaneous fitting with the Dirbe 100$\mu m$ map.
For the 408~MHz map, we also consider the ``desourced version'' (labelled
as Dss, see text for details). 
}
{\tiny
\begin{center}
\begin{tabular}{lccccccccc}
\hline
Template & $\sigma_{Template}$ &C1 & C2 & C3 & WMAP\_K & WMAP\_Ka &WMAP\_Q & WMAP\_V & WMAP\_W \\
\hline
\hline
 \multicolumn{10}{c}{$|b|$$>30^\circ$}  \\
\hline
408 MHz     & $4.88\times 10^5 ~\mu K$ &  $17.0\pm 1.1$ & $12.3\pm 1.2$ & $15.4\pm 2.4$ & $4.7\pm 0.3$  & $2.1\pm 0.3$  & $1.7\pm 0.3$   & $1.1\pm 0.3$ & $0.6\pm 0.3$ \\
408 MHz (Dss)&  $4.79\times 10^5 ~\mu K$ &  $9.3\pm 1.1$ & $8.7\pm 1.2$ & $7.3\pm 2.4$ & $3.7\pm 0.3$  & $2.0\pm 0.3$  & $1.8\pm 0.3$   & $1.4\pm 0.3$ & $1.1\pm 0.3$ \\

1420 MHz    & $2.54\times 10^4 ~\mu K$ &  $20.7\pm 1.1$ & $13.7\pm 1.2$ & $13.2\pm 2.5$ & $5.2\pm 0.3$ & $2.1\pm 0.3$  & $1.4\pm 0.3$   & $0.6\pm 0.3$ & $0.0\pm 0.3$ \\

H$\alpha$   & 0.07~R & 

$2.6\pm 1.1$ & 
$1.4\pm 1.2$ & 
$-2.2\pm 2.4$ & 
$0.1\pm 0.2$ & 
$0.5\pm 0.2$ & 
$0.1\pm 0.2$ &
$0.1\pm 0.2$ &
$0.4\pm 0.2$ \\

\hline
 \multicolumn{10}{c}{$|b|$$>40^\circ$}  \\
\hline
408 MHz     & $4.74\times 10^{5} ~\mu K$ &  $17.0\pm 1.3$ & $13.4\pm 1.4$ & $14.4\pm 2.8$ & $6.1\pm 0.3$ & $2.8\pm 0.3$ & $2.3\pm 0.3$ & $1.6\pm 0.3$ & $1.1\pm 0.3$ \\
408 MHz (Dss)&  $4.71\times 10^5 ~\mu K$ &  $8.0\pm 1.4$ & $8.9\pm 1.4$ & $9.4\pm 2.8$ & $3.8\pm 0.3$  & $2.2\pm 0.3$  & $1.6\pm 0.3$   & $1.4\pm 0.3$ & $0.9\pm 0.3$ \\
1420 MHz    & $2.48\times 10^4~\mu K$ &  $19.1\pm 1.4$ & $13.4\pm 1.5$ & $12.1\pm 2.9$ & $8.1\pm 0.2$ & $2.7\pm 0.3$ & $1.9\pm 0.3$ & $1.2\pm 0.3$ & $0.6\pm 0.3$ \\

H$\alpha$   & 0.06~R & 
$1.1\pm 1.4$ & 
$5.1\pm 1.5$ & 
$1.0\pm 2.9$ & 
$0.3\pm 0.2$ & 
$0.8\pm 0.2$ & 
$0.0\pm 0.2$ &
$0.2\pm 0.2$ &
$0.4\pm 0.2$ \\
\hline
 \multicolumn{10}{c}{$|b|$$>50^\circ$}  \\
\hline
408 MHz     & $4.59\times 10^{5} ~\mu K$ &  $13.8\pm 1.5$ & $9.4\pm 1.6$ & $10.5\pm 3.1$ & $5.8\pm 0.3$ & $3.5\pm 0.3$ & $2.8\pm 0.3$ & $1.9\pm 0.3$ & $1.7\pm 0.4$ \\
408 MHz (Dss)&  $4.69\times 10^5 ~\mu K$ &  $5.5\pm 1.5$ & $5.6\pm 1.6$ & $6.0\pm 3.1$ & $3.4\pm 0.3$  & $2.2\pm 0.3$  & $1.7\pm 0.3$   & $1.2\pm 0.3$ & $1.0\pm 0.4$ \\
1420 MHz    &  $2.45\times 10^4 ~\mu K$ &  $15.3\pm 1.6$ & $9.6\pm 1.6$ & $6.8\pm 3.3$ & $5.8\pm 0.3$ & $3.0\pm 0.3$ & $2.0\pm 0.3$ & $1.1\pm 0.3$ & $0.8\pm 0.4$  \\

H$\alpha$   & 0.05~R & 

$-1.0\pm 1.5$ &
$4.6\pm 1.6$ & 
$-1.1\pm 3.3$ &
$0.7\pm 0.2$ & 
$0.9\pm 0.2$ &
$0.2\pm 0.2$ &
$0.4\pm 0.2$ &
$0.3\pm 0.3$ \\

\hline
\hline
\end{tabular}
\label{tab_templates}
\end{center}
}
\end{table*}

\begin{table*}
\begin{center}
\caption{Same as Table~\ref{tab_templates}, but for correlations with the  
DIRBE 100 and 240 $ \mu$m maps. These numbers were obtained from a simultaneous
fit with the H$\alpha$ map. Units are $\mu$K, except for the rms of the
DIRBE maps ($\sigma_{DIRBE}$) which is given in MJy sr$^{-1}$.}
\label{tab_dirbe}
{\tiny
\begin{tabular}{lccccrrrrr}
\hline 
DIRBE   &  $\sigma_{DIRBE}$  & C1  &  C2 & C3 & WMAP-K & WMAP-Ka & WMAP-Q & WMAP-V  & WMAP-W \\
 \hline 
 \hline 
 \multicolumn{10}{c}{$|b|$$>30^\circ$}  \\
 \hline 
100$\mu$m &  0.11 & 
$7.4\pm 1.1$  & 
$7.5\pm 1.1$  & 
$6.5\pm 2.3$  & 
$2.9\pm 0.2$  &
$0.5\pm 0.2$  & 
$0.0\pm 0.1$  & 
$-0.4\pm 0.2$  & 
$-0.5\pm0.2$   \\

240$\mu$m &  0.27 & 

$6.0\pm  1.1$  & 
$3.4\pm  1.1$  & 
$6.5\pm  2.4$  & 
$2.1\pm  0.2$  & 
$0.3\pm  0.2$  & 
$0.1\pm  0.2$  & 
$-0.4\pm  0.2$  & 
$-0.4 \pm 0.2$   \\
 \hline 
\multicolumn{10}{c}{$|b|$$>40^\circ$}  \\
 \hline 
  100 $\mu$m &  0.07 & 
  $      3.7 \pm       1.4 $ &
$      5.3 \pm     1.4 $  &
$      1.8 \pm       2.8 $ &
$      3.7\pm     0.2 $  & 
$      2.4 \pm      0.2 $  &
$      1.4 \pm      0.2$  &
$      1.4 \pm      0.2 $  &
$      1.3 \pm 0.3$ \\

240$\mu$m  &  0.22 & 
$      3.8 \pm       1.3 $  & 
$    0.7 \pm       1.4$  & 
$      -0.8 \pm       2.9$  & 
$     0.9 \pm     0.2 $  & 
$     0.2 \pm      0.2 $  & 
$     -0.2 \pm      0.2 $  & 
$     -0.5 \pm      0.2$  & 
$     -0.3 \pm    0.2$  \\
 \hline 
\multicolumn{10}{c}{$|b|$$>50^\circ$}  \\
 \hline 
 
 100 $\mu$m &  0.06 & 

$      6.1 \pm       1.5 $  & 
$      5.1 \pm     1.6$  & 
$      -3.1 \pm       3.3 $  & 
$      3.9 \pm     0.2 $  & 
$      2.8 \pm      0.2 $  & 
$      2.2 \pm   0.2 $  & 
$      2.3 \pm      0.2 $  & 
$      2.1 \pm     0.3 $    \\

240$\mu$m &  0.21 & 

$      7.1\pm       1.5 $  & 
$   2.0\pm     1.6 $  & 
$    -5.3 \pm       3.2 $  & 
$     1.3\pm    0.2 $  & 
$     0.6 \pm      0.2 $  & 
$     0.5 \pm    0.2$  & 
$     0.1 \pm      0.2$  & 
$     0.2 \pm     0.3 $    \\   

\hline
\hline 
\end{tabular}
}
\end{center}
\end{table*}

\section{Discussion}

We find a highly significant correlated signal between the COSMOSOMAS and
WMAP maps in the three galactic cuts.  The amplitude of the common signal
ranges between $\sim 30$ and $\sim 40$ $\mu$K depending of the pair of
channels and the galactic cut considered. The amplitude of this signal
decreases as we move from low to high frequency WMAP channels.  This is a
clear indication of the presence at some level of foreground signals with
a negative spectral index. 

The average values found in the correlation between the COSMOSOMAS maps
and the WMAP channel is $29.7\pm 1.0$ $ \mu$K, which corresponds to the
weighted-mean of the values found between WMAP\_V and the three
frequencies of COSMOSOMAS for a galactic cut $|b| >40^\circ$  (the V band was
choosen because it presents the lowest galactic contamination). 
This value is in good agreement with the amplitude of the signal in the filtered
WMAP\_V channel (29.0 $\mu$K at $|b| >40^\circ$), and with the amplitude
of the CMB signal expected in the COSMOSOMAS maps according to simulations
using the best fit $\Lambda$-CDM  model to the WMAP power spectrum 
($\approx 30$ $\mu$K).  This result indicates the realibility of our data and 
the consistency of our technique to build the maps.

Diffuse galactic synchrotron is not expected to be significant over the angular scale to which
COSMOSOMAS is sensitive due to the approximate power law for the galactic  signal in spherical
harmonics of $C_\ell \propto \ell^{-2}$. We can use the upper limit on galactic synchrotron at
10 GHz of $\Delta T_\ell   < 28~\mu$K from the Tenerife results (Guti\'errez et al 2000) at
high galactic latitudes to estimate the expected rms in the C2 map of $<8~\mu K$, assuming a
synchrotron spectral index of -3. This is in very good agreement with the  common signal found
with the 408 MHz 'cleaned map' (see Tables~7 and 9). As we can see in Table~7, the typical 
amplitudes of the correlated signal with the uncleaned 408 MHz map are  higher ($\sim 12-17$ 
$\mu K$). The excess of correlated signal with respect to the desourced map is
mostly independent on galactic latitude, as it is expected for a
random population of extragalactic radiosources. 

Comparing the correlated signal of COSMOSOMAS with the WMAP\_K and WMAP\_V
channels it is possible to  estimate the amplitude of the foreground
contamination in each COSMOSOMAS channel; for instance in the C2 channel
this foreground signal is $22.5\pm 1.7$ $\mu$K at $|b| >30^\circ$ . 
Galactic synchrotron, free-free and DIRBE correlated emission can account for
$\approx 12\mu$K.  The most likely contributor to the remaining foreground
signal ($\approx 19$ $\mu$K) is an extragalactic background of radiosources. 
In fact, although the contribution of the strong radio sources has been
masked, a signal due to extragalactic unresolved radiosources is also
expected.  The amplitude of such contribution can be estimated from the
work by Franceschini et al. (1989), Toffolatti et al. (1998) and, more
recently, de Zotti et al. (2005). This requires the application of the
instrument window function to the angular power spectrum of the radio
sources. Due to the Poissonian distribution of sources the power spectrum
is characterised by a constant $C_\ell$. At 15 GHz for  sources
fainter than 0.7-0.9 Jy the value of $C_\ell$ is expected to be in the range
0.75 - 1.25 $\mu K^2$ from the model of de Zotti et al. respectively (Toffolatti,
private com.). Applying the COSMOSOMAS window function calculated in
section 2.3, we find the range of expected rms in the COSMOSOMAS C2 
map of 20 - 26 $\mu K$, in agreement with the excess signal found. We conclude that apart 
from CMB fluctuations, the signal due to
unresolved radiosources is the main contributor to the common signal
between COSMOSOMAS and the lowest frequency WMAP channels. 

\begin{table}
\caption{Spectral index (in temperature) 
obtained for three galactic cuts from the 
correlation analysis of the 408~MHz 'cleaned map' and the COSMOSOMAS maps.
These numbers were derived from the coupling constants as 
$\alpha = (\nu/408~MHz)^\beta$.}
\begin{center}
\begin{tabular}{cccc}
\hline
\hline
Gal. cut & C1 & C2 & C3 \\
\hline
$|b|>30^{\circ}$ & $-3.16^{+0.03}_{-0.04}$ & $-3.05^{+0.04}_{-0.04}$  & 
$-3.01^{+0.08}_{-0.11}$ \\
\\
$|b|>40^{\circ}$ & $-3.20^{+0.05}_{-0.05}$ & $-3.03^{+0.04}_{-0.05}$  & 
$-2.94^{+0.07}_{-0.10}$ \\
\\
$|b|>50^{\circ}$ & $-3.30^{+0.07}_{-0.09}$ & $-3.16^{+0.07}_{-0.09}$  & 
$-3.05^{+0.11}_{-0.20}$ \\
\hline
\hline 
\end{tabular}
\label{tab_index}
\end{center}
\end{table}

We show in  Table~7 that the correlation between H$\alpha$
map and the COSMOSOMAS channels is below $\sim 5$ $\mu$K. At 12.7GHz the
expected free-free to H$\alpha$ emission ratio is 40 $\mu$K/R for an
ionized medium with 8000 K (Dickinson et al 2004). To estimate the
corresponding correlation signal for a particular region one multiples
this ratio by the rms of the template map in that region. For $|b| >
30^\circ$ the rms in the H$\alpha$ map is 0.07 R, which leads to an
estimated free-free contribution of $\sim 3$ $\mu$K. This is comparable to the 
correlations measured at this galactic cut thus the  amplitude of the
COSMOSOMAS/WMAP correlated signals with H$\alpha$ is compatible with a
free-free spectral index.

We find clear correlated signals between the COSMOSOMAS channels and the
DIRBE maps at $100\mu$m and $240\mu$m.  At $|b|>30^{\circ}$ the weighted-average
amplitudes are $7.3\pm 0.7$ $ \mu$K and  $5.0\pm 0.7$ $\mu$K,
respectively.  The amplitude of the common signal between COSMOSOMAS and
DIRBE maps slightly decreases with galactic latitude. The cross-correlations of 
DIRBE with WMAP also shows a
progressively increasing signal with decreasing frequency with some
evidence for flattening below $\sim 17$  GHz (see Figure~9). In fact, for the
signal correlated with DIRBE 100 $\mu$m the spectral index  between 33 GHz
and 22 GHz is -4.3,  increases to -2.7 between 22 and 16.3 GHz, and becomes
nearly flat at the COSMOSOMAS  frequencies.   This spectral behaviour is
not compatible with the properties expected for thermal dust,
synchrotron or free-free galactic foregrounds. It is
however compatible with predictions for anomalous microwave emission
related to spinning dust (Draine \& Lazarian 1998). In Watson et al 2005
we find that a signal with similar properties is located in a relatively
small area ($\sim 3^{\circ}\times 2^{\circ}$) at $b=-18.5^{\circ}$ located
in the Perseus molecular cloud complex.  To determine whether the DIRBE correlated
signal at high galactic latitudes found in this paper  is associated
with similar sources or corresponds to a diffuse foreground will require
better quality data and observations at lower frequencies.

\section{Conclusions}

The first instrument of the COSMOSOMAS experiment has provided
temperature maps of the sky emission at frequencies 12.7, 14.7 and
16.3 GHz covering $\sim 9000$ square degrees with sensitivities from
50 to 120$\mu$K per beam. Known point-like radio sources with fluxes above $\sim
0.8$ Jy are detected in the three COSMOSOMAS maps. The main results obtained 
in the cross correlation analysis of these maps  with templates of CMB and 
galactic foregrounds are:

\begin{itemize}

\item The correlated signal between the COSMOSOMAS maps and the
highest frequency WMAP\_V  map has an amplitude of 29.7$\pm 1.0$ $\mu$K
($|b|>40^\circ$). This is in agreement with the expected amplitude
of the CMB fluctuations observed through the COSMOSOMAS strategy,
using the best fit  $\Lambda$-CDM model to the WMAP power spectrum.  

\item The amplitude of the COSMOSOMAS-WMAP correlated signal increases as one goes
to lower WMAP frequencies, indicating the presence of a foreground signal
superimposed over the CMB fluctuations.  

\item We find evidence for a contribution of unresolved radiosources to
the COSMOSOMAS maps with an amplitude of $\sim 19 $ $\mu$K at 14.7 GHz which we 
interpret  as the result of a randomly distributed
population of unresolved radiosources.

\item At high galactic latitudes significant correlations for
COSMOSOMAS and the 408 MHz and 1420 MHz galactic templates have been
found. The values of the synchrotron spectral index are between -3.20
and -2.94 in the  408 MHz-16.3 GHz range.

\item The correlation analysis between the COSMOSOMAS/WMAP maps,
and DIRBE channels at 100 and 240 $\mu$m indicates also the presence
of common signals at high galactic latitudes ($|b|>30^\circ$).  The
amplitude of these signals rises with a very steep index from $\sim 40$
GHz up to  $\sim 20$ GHz,  flattening in the range 13-17 GHz with 
an amplitude of  5-7$\mu$K. This behaviour is compatible with
the predictions of spinning dust models. The small correlation ($\sim 3$ $\mu$K) between COSMOSOMAS 
and the H$\alpha$ map rules  out free-free emission as
the  emission mechanism responsible of the dust correlated signal detected in
these regions. 

\end{itemize}

\section*{ACKNOWLEDGEMENTS} 

We would like to thank L. Piccirillo for suggestions in the design of
the experiment. We also thank to L. Toffolatti who kindly
helped us with the estimations of the contribution due to unsolved
radiosources and F. Atrio, F. Villa,  J. Delabrouille, and  G. Patanchon for valuable 
discussions on this experiment. We acknowledge the mechanical and electronic personnel of
the IAC and Teide Observatory who have collaborated in the operation
and maintenance of the experiment. Partial funding was provided by
grant AYA2001-1657 of the Spanish Ministry of Science and Education. C. M. G. was supported
by the {\it Ram\'on y Cajal} programme of the Spanish Science Ministry.

\label{lastpage}
\bibliographystyle{mn2e}

\end{document}